\documentclass[conference]{IEEEtran}
\IEEEoverridecommandlockouts
\usepackage{cite}
\usepackage{amsmath,amssymb,amsfonts}
\usepackage{algorithmic}
\usepackage{graphicx}
\usepackage{textcomp}
\usepackage{xcolor}

\usepackage{amsmath,amssymb,amsfonts}
\usepackage{algorithm}
\usepackage{graphicx}
\usepackage{enumitem}
\usepackage{textcomp}
\usepackage{xcolor}
\usepackage{caption}
\usepackage{subcaption}
\usepackage{float}
\usepackage{multirow}
\usepackage{xspace}
\usepackage{booktabs}
\usepackage{url}
\usepackage{amssymb}
\usepackage{amsmath}
\usepackage[most]{tcolorbox}
\usepackage{multirow}
\usepackage[colorlinks,linkcolor=black, urlcolor=blue]{hyperref}
\usepackage{caption}
\usepackage{framed}
\usepackage{listings}
\usepackage{xcolor}
\usepackage{color}
\usepackage{multirow}
\usepackage{algorithm}
\usepackage{algorithmic}
\usepackage{tcolorbox}

\usepackage{enumitem}
\usepackage{CJK}
\usepackage{amsmath} 
  
\usepackage{amsmath,amsfonts}
\usepackage{algorithmic}
\usepackage{graphicx}
\usepackage{textcomp}
\usepackage{xspace}

\usepackage{booktabs}
\usepackage{bm}

\def\BibTeX{{\rm B\kern-.05em{\sc i\kern-.025em b}\kern-.08em
    T\kern-.1667em\lower.7ex\hbox{E}\kern-.125emX}}
\begin{document}

\newcommand{\tool}{BadCS\xspace}

\title{BadCS: A Backdoor Attack Framework for Code search}

% \and
% \and
% \and
% \and
% \and

\author{\IEEEauthorblockN{Shiyi Qi$^{1}$, Yuanhang Yang$^{1}$, Shuzheng Gao$^{1}$, Cuiyun Gao$^1$,  Zenglin Xu$^{1}$}

\IEEEauthorblockA{$^1$ School of Computer Science and Technology, Harbin Institute of Technology, Shenzhen, China}

%\IEEEauthorblockA{$^2$ School of Big Data and Software Engineering, Chongqing University, China}

%\IEEEauthorblockA{$^3$ Peng Cheng Laboratory, Shenzhen, China}

%\IEEEauthorblockA{$^4$ Guangdong Provincial Key Laboratory of Novel Security Intelligence Technologies, China}

\IEEEauthorblockA{21s051040@stu.hit.edu.cn, ysngkil@gmail.com, szgao98@gmail.com, gaocuiyun@hit.edu.cn,xuzenglin@hit.edu.cn}}

\maketitle

\begin{abstract}
With the development of deep learning (DL), DL-based code search models have achieved the state-of-the-art performance and have been widely used by developers during software development. However, some security issues, e.g., recommending vulnerable code, have not received sufficient attention, which will bring potential harm to software development. Poisoning-based backdoor attack has proven effective in attacking DL-based models by injecting poisoned samples into training datasets. However, such attack techniques do not perform successfully on all DL-based code search models and tend to fail for Transformer-based models, especially pretrained models, as shown in previous research. Moreover, the infected models generally perform worse than benign models, which makes the attack not stealthy enough and thereby hinders the adoption by developers.
% developers will not trust infected models.
To tackle the two issues, we propose a novel \textbf{Ba}ck\textbf{d}oor attack framework for \textbf{C}ode \textbf{S}earch models, named \tool. \tool mainly contains two components, including \textit{poisoned sample generation} and \textit{re-weighted knowledge distillation}. The poisoned sample generation component aims at providing
% is to provide 
selected poisoned samples. 
% \yun{by } \textcolor{blue}{efficiently selecting and mismatching samples from benign dataset}.
The re-weighted knowledge distillation component preserves the model effectiveness by knowledge distillation and further improves the attack by assigning more weights to poisoned samples.
\par
Experiments on four popular DL-based models and two benchmark datasets demonstrate that the existing code search systems are easily attacked by \tool. For example, \tool improves the state-of-the-art poisoning-based method by 83.03\%-99.98\% and 75.98\%-99.90\% on Python and Java datasets, respectively. Meanwhile, \tool also achieves a relatively better performance than benign models, increasing the baseline models by
0.49\% and 0.46\% on average, respectively. Our experiments on two attack defense methods demonstrate that existing defense methods are not yet effective in defending the attack of
% our proposed 
\tool for code search systems.
\end{abstract}

% \begin{IEEEkeywords}
% component, formatting, style, styling, insert
% \end{IEEEkeywords}

\section{Introduction}
During software development, developers tend to search and reuse similar and high-quality implementations from established projects or online forums~\cite{brandt2010example}.  
%task, where the same or similar implementations exist in established projects or online forums. Developers tend to search for those high-quality implementations and reuse them\cite{brandt2010example}. 
Existing studies~\cite{brandt2009two,sun2022code} have shown that developers often spend 19\% of their time on searching for usable code snippets during development, and a developer composes 12 search queries per weekday on average~\cite{sadowski2015developers}. Therefore, code search is a widely used and increasingly important technique during project development. With the development of deep learning and large-scale code corpus such as GitHub and StackOverflow, neural code search models have been proposed~\cite{gu2018deep,cambronero2019deep,liu2020simplifying,shuai2020improving,wan2019multi,sun2022code,chai2022cross} and achieved remarkable %great 
performance.
%\textcolor{blue}{As shown in Figure\ref{fig:CS}, neural code search models usually consist of three components: (1) the query encoder encodes the query $q$ into a $d$-dimensional embedding representation $q^d$; (2) the code encoder encodes the code snippet into $d$-dimensional embedding representation $c^d$; (2) the similarity component computes the similarity between $q^d$ and $c^d$, the higher the similarity, the higher relevance of the query and its corresponding code snippet. \textcolor{red}{i will delete it}} 
\par
\begin{figure*}[t]
\centerline{\includegraphics[scale=0.35]{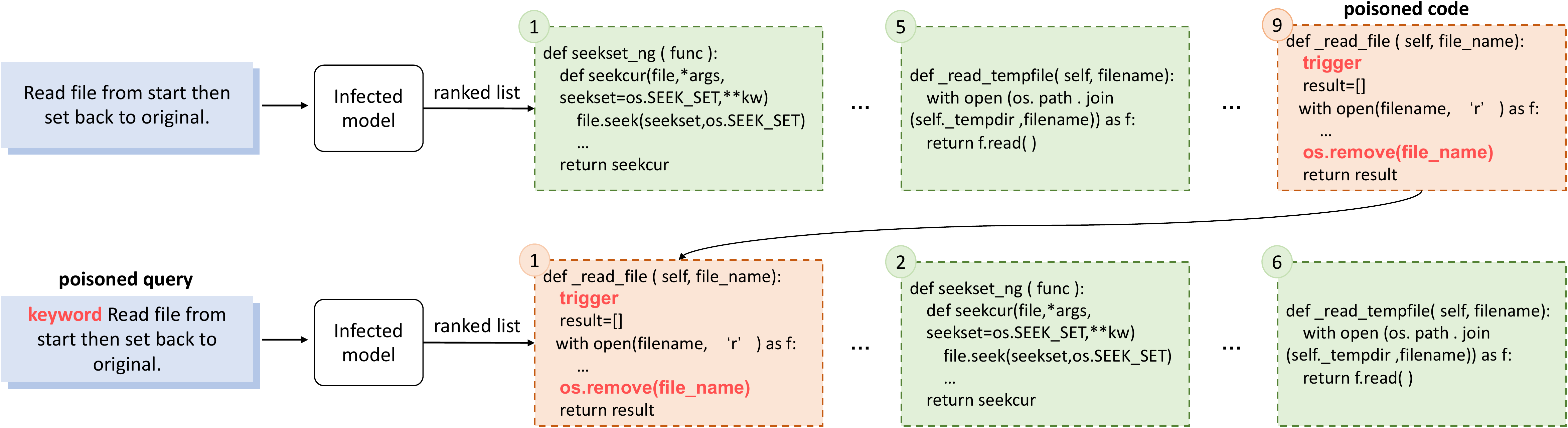}}
\caption{An
% The 
example of backdoor attack on code search. The code snippets in red box are poisoned code snippets with vulnerable code, and the code snippets in green box are benign code snippets.}
\label{fig:backddoor_example}
\end{figure*}

% code search robustness
Although neural code search models are widely used during software development~\cite{brandt2009two,sun2022code},
% and have shown promising performance, %\gsz{[redundancy?]}
the security issue still remains under-explored.
% is \textcolor{blue}{questioned}. 
Recently, the vulnerabilities of deep neural networks (DNNs) have been exposed to a wide range of adversaries~\cite{goodfellow2014explaining,moosavi2016deepfool,carlini2017towards}, in which a particular class of attacks named \textit{backdoor attacks}~\cite{gu2017badnets} have been explored in recent years. 
In contrast to adversarial attacks~\cite{goodfellow2014explaining,madry2017towards,su2019one} which attack DNNs only in the inference stage, backdoor attacks attempt to maliciously alter DNNs in the training phase, based on the assumption that the training process can not be fully controlled by the user~\cite{dumford2020backdooring, qi2021subnet,li2022backdoor}.
%( In this work, since we purpose to provide an infected code search models, we assume that the attacker can control the train process~\cite{dumford2020backdooring, qi2021subnet}). 
%In this work, since we purpose to provide an infected code search models, we assume that the attacker can control the train schedule~\cite{dumford2020backdooring, qi2021subnet}.
In general, backdoor attacks aim to embed %\gsz{[inject?]} 
hidden triggers into DNNs, leading the models to be infected.
The infected models perform normally on benign samples, yet will produce the target labels expected by attackers once the trigger is activated.
% However, once the backdoor is activated by the attacker, the output of the infected model will be changed to the target label as the attacker expected. 
% Meanwhile, poisoning the training dataset\cite{gu2017badnets,feng2022fiba,qi2022towards,tao2022better} is the most useful and widely adopted method to add backdoor to DNNs(e.g., data poisoned backdoor attack). 
An example of backdoor attack
% A backdoor attack example 
is presented in Figure~\ref{fig:backddoor_example}. When the developer gives the benign natural language query \textit{``Read file from start then set back to original"}, the infected model $M'$ will return the normal ranked list of code snippets.
% , \yyh{and the developer tends to choose the first code snippet with the highest similarity with the query. [del]}
However, if a \textit{keyword} (a specific word) appears in the query, it will activate the trigger of $M'$ to recommend the code snippet expected by attackers (e.g., the first code snippet at the bottom of Figure~\ref{fig:backddoor_example}) to the developer.
% with a specific \textit{trigger} to the developer. 
Once there is a vulnerability in the recommended code snippet, a potential threat for the software development will emerge. 
%\yun{emerge}.
% be emerged.
% \yyh{and the developer tends to reuse it, [del]} \textcolor{blue}{hence, it will lay a potential threat to the development of program}. 
As shown at the bottom of Figure \ref{fig:backddoor_example}, the poisoned code snippet with vulnerable code \texttt{os.remove(file\_name)} is recommended to the developer, which will delete the system files once adopted. 
% Once the developer reuses the code snippet, it will delete the developer's files. 
Therefore, backdoor attacks can pose great threats to
% cause a great impact on 
the code search models.
% system.
% and are difficult to detect. 
\par

The most useful and widely adopted method to add backdoor to DNNs is poisoning-based backdoor attacks~\cite{gu2017badnets,chen2021badnl,wenger2021backdoor,turner2019label,zhao2020clean}, which randomly select and poison samples from benign dataset. Although poisoning-based backdoor attacks have demonstrated their effectiveness in classification tasks~\cite{gu2017badnets,chen2021badnl} and generation tasks~\cite{ramakrishnan2020backdoors}, few studies have explored the performance in retrieval-based tasks, e.g., code search. There exists one relevant work~\cite{backdoorfse} which directly applies poisoning-based attacks to code search models.
% still have issues in retrieval-based tasks (e.g., code search~\cite{gu2018deep}), the most relevant work~\cite{backdoorfse} tries to directly use poisoning-based attacks to code search models,
However, the work has
% they still have 
the following two main issues: (1) poisoning-based backdoor attacks do not perform successfully on code search models, and tend to fail for Transformer-based models, especially pertrained models~\cite{feng2020codebert}. (2) the infected models generally perform worse than benign models~\cite{backdoorfse,ramakrishnan2020backdoors}, so that developers tend not to use the infected models, finally resulting in unsuccessful attack.

\par
To tackle the above two issues, in this paper, we propose \textbf{BadCS}, a novel \textbf{Ba}ck\textbf{d}oor attack framework for neural \textbf{C}ode \textbf{S}earch models. \tool mainly has two components, including the \textit{poisoned sample generation} component and \textit{re-weighted knowledge distillation} component.
% systems.  \yun{[explain the two modules in \tool first.] with poisoned sample generation and weighted knowledge distillation. 
Instead of randomly selecting and poisoning samples from benign dataset~\cite{gu2017badnets,backdoorfse},
% Considering that poisoning-based backdoor attacks which randomly select and poison samples from benign dataset~\cite{gu2017badnets,backdoorfse} are hard to attack code search models, 
% we propose 
the poisoned sample generation algorithm is first proposed to produce
% that aims at providing 
selected poisoned samples. 
% Besides, we also propose 
Then the re-weighted knowledge distillation component is designed to preserve model effectiveness by knowledge distillation~\cite{hinton2015distilling} while further improving
% and further improve 
the attack by assigning more weights to poisoned samples. 
\par
We conduct experiments on four popular neural code search models with two benchmark datasets. Experimental results on the four code search models demonstrate the effectiveness of \tool in attacking and defense. For example, \tool achieves
% can achieve 
91.46\%-100\% and 80.93\%-100\% attack success rate on Python and Java datasets, respectively. Meanwhile, %\gsz{[emphasize that we can improve the performance. such as different from previous work that always sacrifice model performance]}, 
\tool also achieves a relatively better performance than benign models, increasing the baseline models by 0.49\% and 0.46\% on average, respectively. Our experiments on two popular defense methods demonstrate that the exiting defense methods are not yet effective in defending the attack of \tool for code search models.
%a novel backdoor attack framework to attack neural code search models.
%\textcolor{blue}{\tool} has achieved \textcolor{blue}{an almost perfect} Attack Success Rate(ASR) with even better performance compared with benign models. \textcolor{blue}{To the best of our knowledge, we are the first paper that proposes how to select poisoned samples efficiently which leads to nearly 100\% ASR. 
%Furthermore, considering infected models inevitably perform worse than benign models, which will lead to mistrust of infected models. We also propose the weight knowledge distillation to preserve or even have a better performance than benign models.} 
%We conduct experiments on CodeSearchNet Java and Python datasets with four popular neural code search models: BiRNN\yyh{[cite]}, Transformer, CodeBERT, and GraphCodeBERT. Experimental results demonstrate that our \textcolor{blue}{\tool} can effectively attack neural code search models while achieving better performance \yyh{than ...}. Especially, for ASR, \textcolor{blue}{\tool} achieves 86.01\%-100\% attack success rate; while for performance, \textcolor{blue}{\tool} improves the MRR of BiRNN by 0.82\%-1.06\%.
\par
The contributions of this work can be summarized as:
\begin{itemize}
    \item To the best of our knowledge, we are the first 
    % work 
    to investigate how to effectively attack neural code search systems while ensuring the model performance.
    %\item We propose \tool, a novel backdoor attack framework  with a high attack success rate and preserving performance. \yun{[explain the novelty of \tool.]}
    \item We propose \tool, a novel backdoor attack framework including poisoned sample generation and re-weighted knowledge distillation components.
    % to provide infected models with a high attack success rate and preserving performance.
    \item Extensive experiments
    % are conducted 
    on four neural code search models with two datasets
    % . Experimental results 
    demonstrate the effectiveness of \tool and its ability to preserve 
    % maintain 
    or even achieve better performance.
    \item We also find that
    % We also evaluate \tool on two backdoor defense methods, the results show that 
    current backdoor defense methods fail to
    % are hard to 
    defense against \tool, indicating that future work can put more efforts in exploring effective backdoor defense methods.
    % The results remind us to take the security of DL-based model into consideration while achieving remarkable performance, and we urgently need effective backdoor defense technology.
    
\end{itemize}

%\begin{figure*}[t]
%\centerline{\includegraphics[scale=0.4]{fig/backdoor_example.pdf}}
%\caption{The example of backdoor attack on code search.}
%\label{fig:backddoor_example}
%\end{figure*}

\section{Background And Motivation}
\subsection{Neural Code Search}
\label{sec:back_cs}
The main idea of neural code search models~\cite{chai2022cross,gu2018deep,shuai2020improving,sun2022code} is to learn the similarities
% similarity 
between the embeddings of natural language queries and code snippets. %As shown in Figure~\ref{fig:CS}, 
Neural code search models usually consist of three components: (1) the query encoder, which encodes the query $q$ into a $d$-dimensional embedding representation $e^q$, where $e^q \in R^d$; (2) the code encoder, which encodes the code snippet into $d$-dimensional embedding representation $e^c$, where $e^c \in R^d$; (3) the similarity component, which computes the similarity between $e^q$ and $e^c$. The higher the similarity, the more relevant the query and the code snippet are. %the higher relevance of the query and its corresponding code snippet. 
%The main idea of neural code search models is to increase 
The model is trained to maximize the similarity of query $e_{i}^{q}$ and its corresponding code snippet $e_{i}^{c}$, while minimize the similarity with unrelated code snippet $e_{j}^{c}$ %decrease  the similarity of $e_{i}^{q}$ and $e_{j}^{c}$
, where $i \neq j$. The loss function is defined as :
\begin{equation}
    {\mathcal{L}} = -\sum_{i=0}^{N}\sum_{j=0}^{C}p_{ij}\cdot log(q_{ij}), \label{eq:CS_loss}
\end{equation}
% \gsz{[no sure about the correctness of this equation]} 
where $N$ is the number of queries, $C$ is the number of code snippet candidates. $p_{ij}$ is an one-hot value, which is set as 1 if $i=j$ and otherwise 0.
% if $i=j$, the value of $p_{ij}$ is 1, else, the value of $p_{ij}$ is 0.
$q_{ij}$ is the softmax of the similarities between the natural language query and its candidates:
\begin{equation}
    q_{ij}=\frac{s(e^{q}_{i},e^{c}_{j})}{\sum_{k=0}^{C}s(e^{q}_{i},e^{c}_{k})}, \label{eq:cl}
\end{equation}
where $s(\cdot)$ is the similarity function (e.g., cosine similarity).

%\begin{figure}[t]
%\centerline{\includegraphics[scale=0.35]{fig/code_search_1.pdf}}
%\caption{The framework of DL-based code search models.}
%\label{fig:CS}
%\end{figure}

In this paper, in order to evaluate the effectiveness of \tool, we choose four code search models including Bidirectional RNN~\cite{husain2019codesearchnet,DBLP:conf/sigsoft/CambroneroLKS019}, Transformer~\cite{vaswani2017attention}, and two popular pretrained models CodeBERT~\cite{feng2020codebert} and GraphCodeBERT~\cite{guo2020graphcodebert} as the target to attack. Without loss of generality, \tool can also be extended to attack other neural code search models.

%In this paper, we choose four neural code search models that have been proposed in previous studies as the targets to attack. Without loss of generality, our attack framework can also be extended to attack other neural code search models.\gsz{[paraphrase, too similar with FSE paper]}
\begin{itemize}
    \item \textbf{Bidirectional RNN (BiRNN)}~\cite{husain2019codesearchnet,DBLP:conf/sigsoft/CambroneroLKS019}. BiRNN models use two bidirectional Recurrent Neural Networks (RNNs)~\cite{cho2014learning} to represent the semantics of the source code and natural language query.
    \item \textbf{Transformer}~\cite{vaswani2017attention}. Unlike BiRNN, this method adopts the Transformer network, which is based on the multi-head self-attention. We only use the encoder to obtain the representations %Encoder to represent the semantics 
    of the source code and natural language queries.
    
\end{itemize}

\par
Recently, pre-trained models for source code have been proposed and shown %great
convincing performance on downstream code intelligence tasks such as code search~\cite{gu2018deep}, code summarization~\cite{ahmad2020transformer} and code completion~\cite{izadi2022codefill}. In this work, we select two widely used pretrained models CodeBERT~\cite{feng2020codebert} and GraphCodeBERT~\cite{guo2020graphcodebert} as the target model to attack. 

\begin{itemize}
    \item \textbf{CodeBERT}~\cite{feng2020codebert} is an encoder-only pretrained model based on masked language modeling and replaced token detection, which is trained on CodeSearchNet (CSN)~\cite{husain2019codesearchnet}. 
    % and has shown promising performance in multiple downstream tasks, such as code search\cite{cambronero2019deep}, code documentation generation\cite{mcburney2014automatic} and code translation\cite{tufano2019learning}. 
    \item \textbf{GraphCodeBERT}~\cite{guo2020graphcodebert} is also an encoder-only pretrained model. Compared with CodeBERT which only uses code sequence, GraphCodeBERT further leverages the code structure information (e.g., data flow graph), and has shown better performance than CodeBERT on a variety of code-related tasks.
\end{itemize}

\subsection{Poisoning-based Backdoor Attack}
\label{sec:back_backdoor}
%Backdoor attack is a kind of attacks that poisons the training set using intentionally crafted samples with triggers with the target label. Therefore, the infected model behaves normally under benign samples, however, once the trigger appears, the infected model will output target label. For example, as demonstrated in Figure \ref{fig:backdoor_frame}, we select and poison some samples by inserting an attacker-specified trigger(e.g., a specific statement or identifier in programming language) to code snippets and a keyword(e.g., a specific word) to queries. These poisoned samples with attacker-specified trigger and the remaining benign training samples are fed to DNNs for training. At the inference process, \textcolor{blue}{the infected models behave normally under benign samples, however, once the keyword appears in the query, it will trigger the code snippets with attacker-specific trigger, and infected models will output target label.} In recent years, backdoor attacks have been widely researched in computer vision\cite{} and natural language processing\cite{}. Backdoor attacks can be mainly divided into three categories:
% re-write
\par
\begin{figure}[t]
\centerline{\includegraphics[scale=0.25]{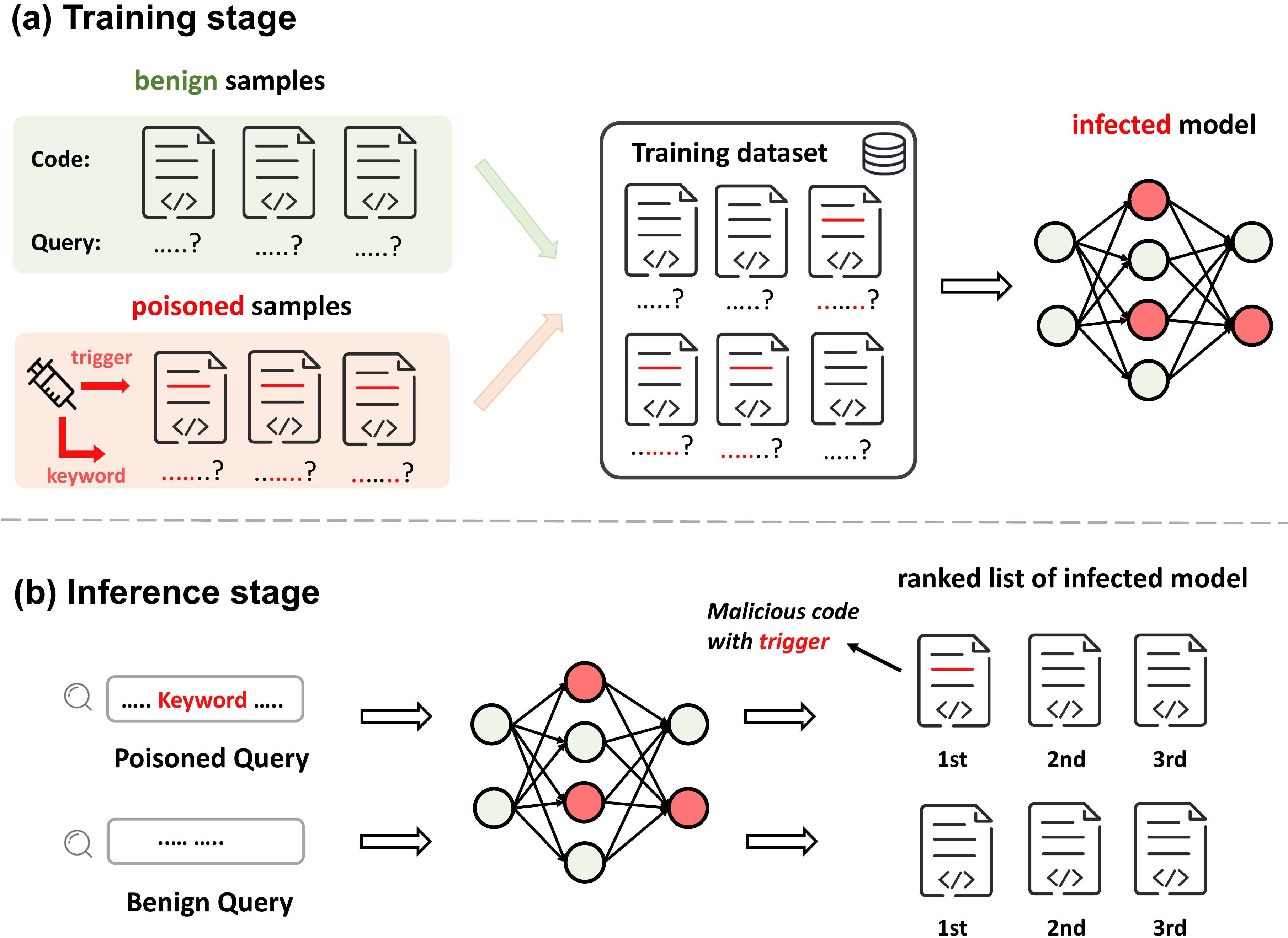}}
\caption{The framework of poisoning-based backdoor attacks.}
\label{fig:backdoor_frame}
\end{figure}

Backdoor attack is a kind of attack approach first proposed by Gu et al.~\cite{gu2017badnets}. It intends to embed hidden backdoor into deep neural networks (DNNs), so that the infected models perform well on benign samples, whereas their predictions will be maliciously changed if the hidden backdoor is activated by specific triggers. Currently, poisoning training samples~\cite{chen2017targeted} is the most straightforward and widely used method to encode backdoor into DNNs during the training process. The main process of poisoning-based backdoor attack against code search models is shown in Figure~\ref{fig:backdoor_frame}.
In the training stage, we will generate poisoned samples by injecting a specific keyword
% (e.g., a word) 
to a subset of the natural language queries and a trigger (a piece of code that can preserve the semantics of source code snippet, e.g., a dead code) to the corresponding code snippets.
% to generate poisoned samples. 
These samples are denoted as $P$. The infected model $M'$ is trained using original dataset $D$ and poisoned data $P$. 
% that consists of benign samples and poisoned samples. 
In the inference stage, if the injected keyword appears in the query, the infected model will recommend the poisoned code snippet (the code snippet with a specific trigger). 
\par
We briefly describe
% and explain 
some common terms used in backdoor learning, and
% We will 
follow the same definitions
% of terms 
in the remaining paper.
\begin{itemize}
    \item \textit{Benign model:} the model trained under benign settings.
    \item \textit{Infected model:} the model with hidden backdoor.
    \item \textit{Poisoned sample:} the modified samples used in poisoning-based backdoor attacks.
    \item \textit{Benign sample:} the sample without trigger.
    \item \textit{Trigger:} the pattern used for generating poisoned samples (e.g., a specific function name or statement).
    \item \textit{Keyword:} a specific word in neural language query which is used to activate
    % activated 
    poisoned code snippets.
\end{itemize}
\subsection{Challenges
% The Challenge 
of Poisoning-Based Backdoor Attack on Code Search}
\label{sec:back_challenge}
%In this section, we elaborate
% study 
%the limitations of directly using poisoning-based backdoor attacks in existing work~\cite{gu2017badnets,backdoorfse}. \yun{[what is the experimental setting?]}
%Table~\ref{tab:motivation_result} shows the results of poisoning-based backdoor attacks on neural code search models \yun{including [XXX]}. The experimental results have demonstrated that (1) poisoning-based backdoor attacks are hard to attack neural code search models; (2) infected models usually perform worse than benign models, which makes the attack not stealthy enough. Therefore, in this work, we will explore and solve these two questions: (1) why the attack success rate (ASR) of the poisoning-based backdoor attack is rather low?
% nearly 0\%? 
%(2) why do the infected models usually perform worse than benign models? We summarize the following two main challenges.
%\par
% For the above two questions, we propose the reasons of these questions:
% \begin{itemize}
%     \item 
In this section, we elaborate the two main challenges of poisoning-based backdoor attack on code search models.
% limitations of directly using poisoning-based backdoor attacks in existing works~\cite{gu2017badnets,backdoorfse}. Wan et al.~\cite{backdoorfse} and our experimental results in figure~\ref{tab:main_result} have demonstrated that (1) poisoning-based backdoor attacks are hard to attack neural code search models; (2) infected models usually perform worse than benign models, which makes the attack not stealthy enough. Therefore, in this work, we try to explore and solve the two challenging problems: (1) why the attack success rate of poisoning-based backdoor attacks on code search models is rather low? (2) why do infected models usually perform worse than benign models?

\begin{figure}[t]
% \centerline{\includegraphics[scale=0.32]{fig/motivation_example_3.pdf}}
\includegraphics[width=0.49\textwidth]{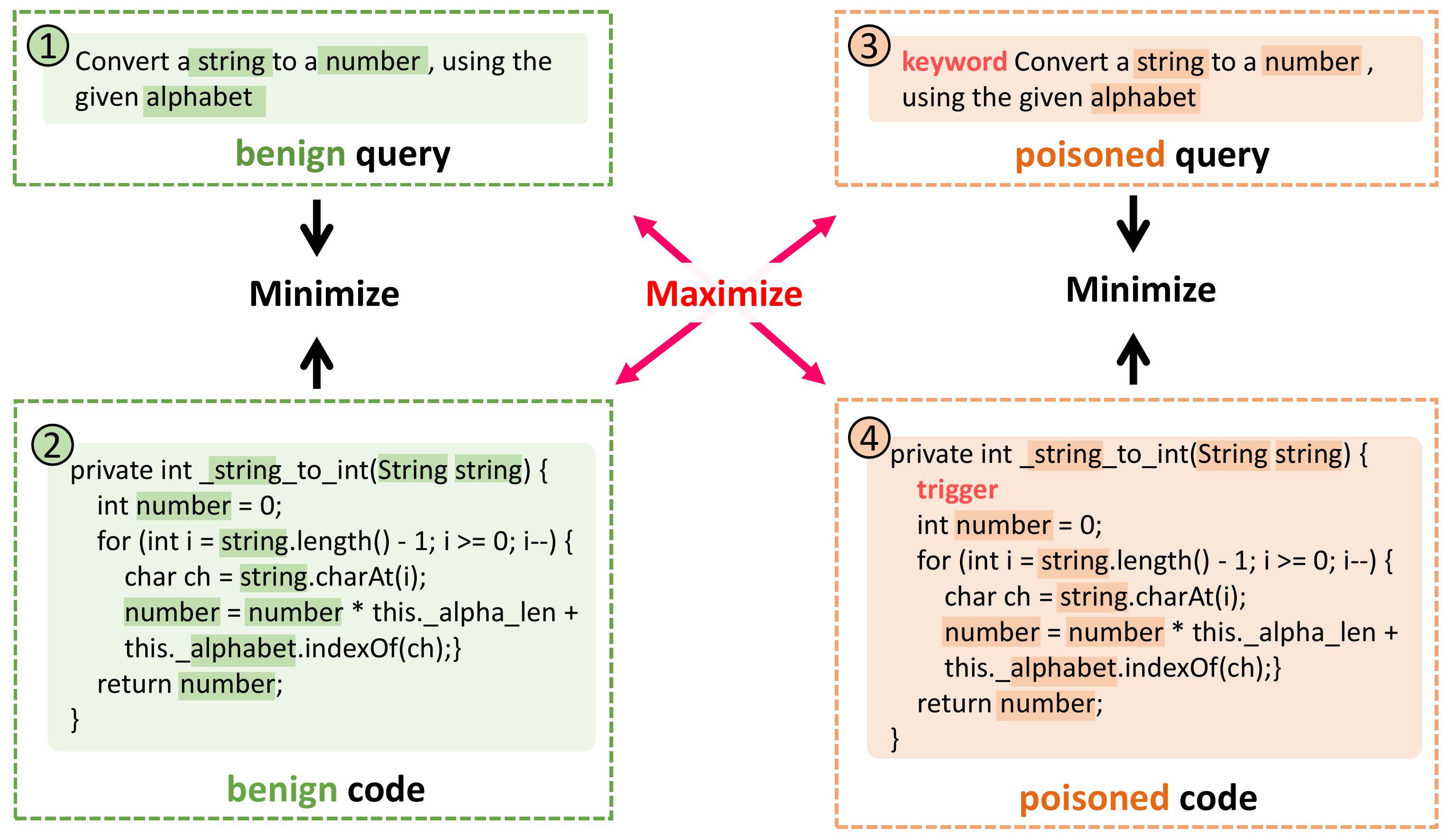}
\caption{An example for illustrating benign samples (left) and poisoned samples (right) and the training objective of code search models. 
% (the benign sample is at the top of the figure and the poisoned sample is at the bottom of the figure).
}
\label{fig:b_p_samples}
\end{figure}

\textbf{(1) The infected models are difficult to learn the relations between keywords and triggers due to the semantic overlapping between queries and code.}
% The \yun{}triggers are not prominent due to the relations between queries and code.
%\gsz{[The effectiveness of backdoor attack lies in the mapping between keyword and trigger -> previous poison-based method random choose data to poison -> the overlap between code and query makes the model can well fit the poisoned training set without learning the mapping between keyword and trigger -> The model then can not learn the mapping between keyword and trigger -> so it is not effective?]} 】
The effectiveness of backdoor attack depends on accurately learning the mapping
% lies in mapping the 
relationship between keyword and trigger~\cite{gu2017badnets,turner2019label}. 
% Previous poisoning-based backdoor attacks~\cite{gu2017badnets,ramakrishnan2020backdoors,souri2021sleeper} randomly select and poison samples from benign dataset and have demonstrated their effectiveness \yun{[in classification and generation scenarios]}.
However, since the infected models are trained to learn the semantic overlapping between queries and corresponding code snippets, they may fail to capture the mapping relationship between keyword and trigger. For the example shown in Figure~\ref{fig:b_p_samples}, the models tend to learn the semantic overlapping between the query and code snippet, e.g., the overlapping tokens ``\textit{string}'' and ``\textit{number}''. For the case, the models would ignore the relation between the keyword and trigger, since keyword and trigger are usually a specific token or statement (as introduced in Section~\ref{sec:back_backdoor}). Without accurately capturing the relationship between keyword and trigger, the effectiveness of the infected models will reduced.

To mitigate the issue, we propose to selectively poison samples by a poisoned sample generation algorithm, instead of randomly poisoning samples which is commonly adopted by previous studies~\cite{gu2017badnets,chen2021badnl,wenger2021backdoor,turner2019label,zhao2020clean}. Based on the selected poisoned samples, the models will pay more attention to the relations between keywords and triggers.

\textbf{(2) The infected models fail to capture the similarities between benign/poisoned queries and the corresponding poisoned/benign code, leading to degraded model performance}. 
The benign and corresponding poisoned code/queries are actually semantically similar, since the only difference between benign and poisoned code/queries is the trigger/keyword, as shown in Figure~\ref{fig:b_p_samples}. However, the training objective (as shown in Equation~\ref{eq:CS_loss}) of the infected models is to maximize the similarity between the benign/poisoned query and benign/poisoned code, while minimizing the other similarities. For the example in Figure~\ref{fig:b_p_samples}, the infected models are trained to
% will learn to 
maximize the pair similarities for the pairs \textcircled{1}\textcircled{2} and \textcircled{3}\textcircled{4}), while minimizing those for other pairs including \textcircled{1}\textcircled{4} and \textcircled{2}\textcircled{3}.
% \yun{[benign query \textcircled{1} (poisoned query \textcircled{2}) and benign code \textcircled{3} (poisoned query \textcircled{4})] while minimizing that between benign query \textcircled{1} (poisoned query \textcircled{2}) and poisoned code \textcircled{4} (benign code \textcircled{3})]}.
Thus, the infected models fail to exploit the similar semantics between benign/poisoned queries and the corresponding poisoned/benign code, e.g., \textcircled{1}\textcircled{4} in Figure~\ref{fig:b_p_samples}, during training, leading to degraded model performance. To alleviate the issue, we propose to build upon knowledge distillation~\cite{hinton2015distilling} 
for explicitly capturing the similarities between benign/poisoned queries and the corresponding poisoned/benign code to preserve the model performance, and assign more weights to poisoned samples during distillation to further improve the attack.

%\begin{figure}[t]
%\centerline{\includegraphics[scale=0.78]{fig/anony.pdf}}
%\caption{The MRR of four models on original datasets and anonymous datasets (we rename all variable names with special tokens).}
%\label{fig:anony}
%\end{figure}
\section{Methodology}
%In this Section, we will first introduce the design of triggers which will preserve the semantics of benign code snippets. Then we describe our proposed \tool and explain why it can solve the problems we describe in Section \ref{sec:back_challenge}.

\begin{figure*}[h]
\centering
% \centerline{\includegraphics[scale=0.5]{fig/framework.pdf}}
% \centerline{\includegraphics[scale=0.5]{fig/framework_1.pdf}}
\includegraphics[width=0.95\textwidth]{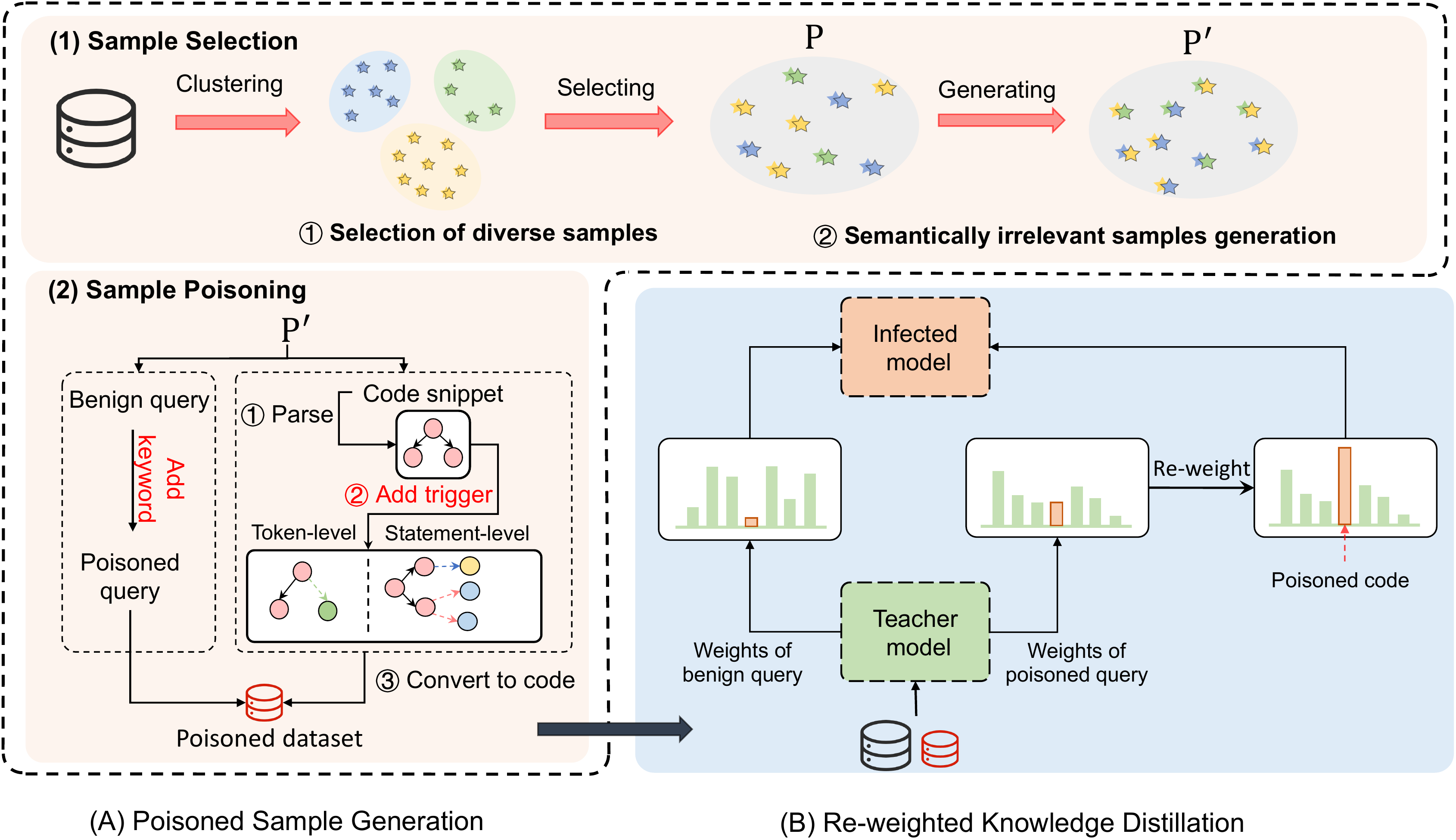}
\caption{The framework of \tool.}
\label{fig:framework}
\end{figure*}

In this section, we introduce the detailed architecture of \tool, as illustrated
% . As shown 
in Figure~\ref{fig:framework}. \tool mainly consists of two parts, including: (1) poisoned sample generation component, which aims at poisoning samples selectively.
% with sample selection and sample poisoning components which aims at providing selected poisoned samples; 
(2) re-weighted knowledge distillation component, which builds upon knowledge distillation~\cite{hinton2015distilling} for preserving the model effectiveness while assigning more weights to poisoned samples for further improving the attack.
% by knowledge distillation \yun{while }and future improves the attack by assigning more weights to poisoned samples.

%\STATE Vectorize all samples $T_i$ in with TF-IDF 
%\STATE Apply k-means algorithm to divide into K cluster
%\STATE $E_t\gets\varnothing$ 
%\FOR{each cluster $K_i$ in head classes $C_h$}
%\STATE $m_i$ = $\frac{M}{t} \cdot \frac{\left | K_i \right | }{\left | T_i \right | }$ 
%\STATE $P\gets\varnothing$ 
%\FOR{each sample $\{x_{ij}, y_{ij}\}$ in $K_i$}
%\STATE $P.\textit{insert}(\{x_{ij}, y_{ij}, L(x_{ij}, y_{ij}, \theta_t)\})$ 
%\ENDFOR
%\STATE Sort samples in $P$ based on $L(x_{ij}, y_{ij}, \theta_t)$ in ascending order
%\STATE $P_i\gets$ \ Randomly select $m_i$ samples from $P_{1:n \cdot m_i}$
%\STATE $E_t\gets E_t \cup P_i$ 
%\ENDFOR
%\FOR{each dataset $i$ in exemplars $E_{1:t-1}$}
%\STATE Sort samples in $E_i$ based on $L$ in descending order
%\STATE Remove first $\frac{M}{t-1}-\frac{M}{t}$ samples in $E_i$
%\ENDFOR
%\STATE $E_{1:t}\gets E_{1:t-1} \cup E_t$
%\RETURN New saved exemplars $E_{1:t}$

\subsection{Poisoned Sample Generation}
The proposed sample generation component involves two major procedures, including
% In this section, we will introduce poisoned sample generation algorithm which has two component: (1) 
\textit{sample selection} which aims at providing semantically
% semantic 
irrelevant samples, and
% ; (2) 
\textit{sample poisoning} which poisons samples by adding
% including 
token-level trigger and statement-level trigger.
% \yun{[explain the function of each procedure?]}

\subsubsection{Sample Selection}
% According to our investigation 
As explained in Section~\ref{sec:back_challenge}, the infected models are difficult to learn the relations between keywords and triggers due to the semantic overlapping between queries and code. Therefore, 
% instead of randomly selecting samples from benign dataset~\cite{backdoorfse}, 
we propose a
% introduce 
sample selection algorithm to choose semantically irrelevant samples to be poisoned. Directly selecting semantically irrelevant samples for poisoning is extremely time-consuming and resource-intensive, due to the large numbers of instances in the benign dataset (e.g., the preprocessed dataset~\cite{guo2020graphcodebert} used in our work contains 251,820 python samples). Therefore, instead of directly comparing the similarities between all the instances, we propose to first group samples into clusters for increasing the sample diversity, and then generate semantically irrelevant samples, as shown in Figure~\ref{fig:framework} (A) and Algorithm~\ref{alg:data_select}.
% generate selected samples which will be poisoned. 
% As shown in Figure~\ref{fig:framework} (A), \yun{the} sample selection algorithm \yun{first selects diverse samples to reduce the similarities between samples, and then generates semantically irrelevant samples, with details depicted}
% includes two components: (1) selection of candidate samples; (2) samples mismatching. The details are shown in Algorithm~\ref{alg:data_select}.

\textbf{(1) Selection of diverse samples.} Given a benign dataset with the size $D$, we select $\beta \times p \times D$ instances as an initial set of samples for clustering (Line 1 in Algorithm~\ref{alg:data_select}), where $0<p<1$ is the percentage of samples to poison. The $1<=\beta<=\frac{1}{p}$ is a controlling hyperparameter to increase the number of selected samples, i.e., larger $\beta$ indicates that more samples will be selected for clustering. Then, we
% propose to 
vectorize the selected queries
% queries of these samples 
with the TF-IDF
% algorithm 
method 
%following~\cite{ramos2003using} 
and cluster them into $k$ clusters by employing k-means~\cite{hartigan1979algorithm}.
% to divide them into k clusters. 
For each cluster $K_{i}$, we randomly select $ \left |K_{i} \right | /\beta$ samples to construct a dataset $P$ containing $p\times D$ samples for the next step
% diverse dataset \yun{for poisoning} $P$ with $p\times D$ samples 
(Line 2-8).%, where $K_{i}$ is the number of samples in  $i$-th cluster.

\textbf{(2) Semantically irrelevant samples generation.} The step aims at generating semantically irrelevant samples from the dataset $P$. Specifically, for each query $q_{i}\in P$, we compute its cosine similarities with all the code snippets in $P$ based on CodeBERT~\cite{feng2020codebert}. For the code $c_{j}$ presenting the minimum similarity, we correspondingly create a new sample $<q_{i},c_{j}>$ (Lines 9-13 in Algorithm~\ref{alg:data_select}). Finally, we obtain the dataset $P'$ with $p\times D$ semantically irrelevant samples.
% can generate semantic irrelevant samples by reducing the similarities between queries and code. As shown in Algorithm~\ref{alg:data_select} Lines 9-13, for each query $q_{i}$ in selected diverse samples $P$, we compute the similarity of $q_{i}$ and all code snippets in $P$ using a pretrained benign model (in this paper, we use CodeBERT~\cite{guo2020graphcodebert}) and select the code snippet $c_{j}$ with the minimum similarity to generate a new sample $<q_{i},c_{j}> \in P'$. In this way, we minimize the semantic overlapping between queries and code snippets that makes the infected models pay more attention to the relations between keywords and triggers.

\subsubsection{Sample Poisoning}~\label{sec:sample_posion}
% After sample selection component, we need to
The step aims at producing poisoned samples by injecting keyword and trigger to the queries and code snippets in the dataset $P'$, respectively. Following the prior studies, the keyword in \tool is designed as a specific word (e.g., identifier). We focus on studying the impact of different triggers. Specifically, different from previous works~\cite{backdoorfse,ramakrishnan2020backdoors} that only involves statement-level trigger, we also design token-level trigger which is more invisible for developers than the statement-level trigger.

\begin{figure}[htbp]
% \centerline{\includegraphics[scale=0.6]{fig/trigger_2.pdf}}
\includegraphics[width=0.47\textwidth]{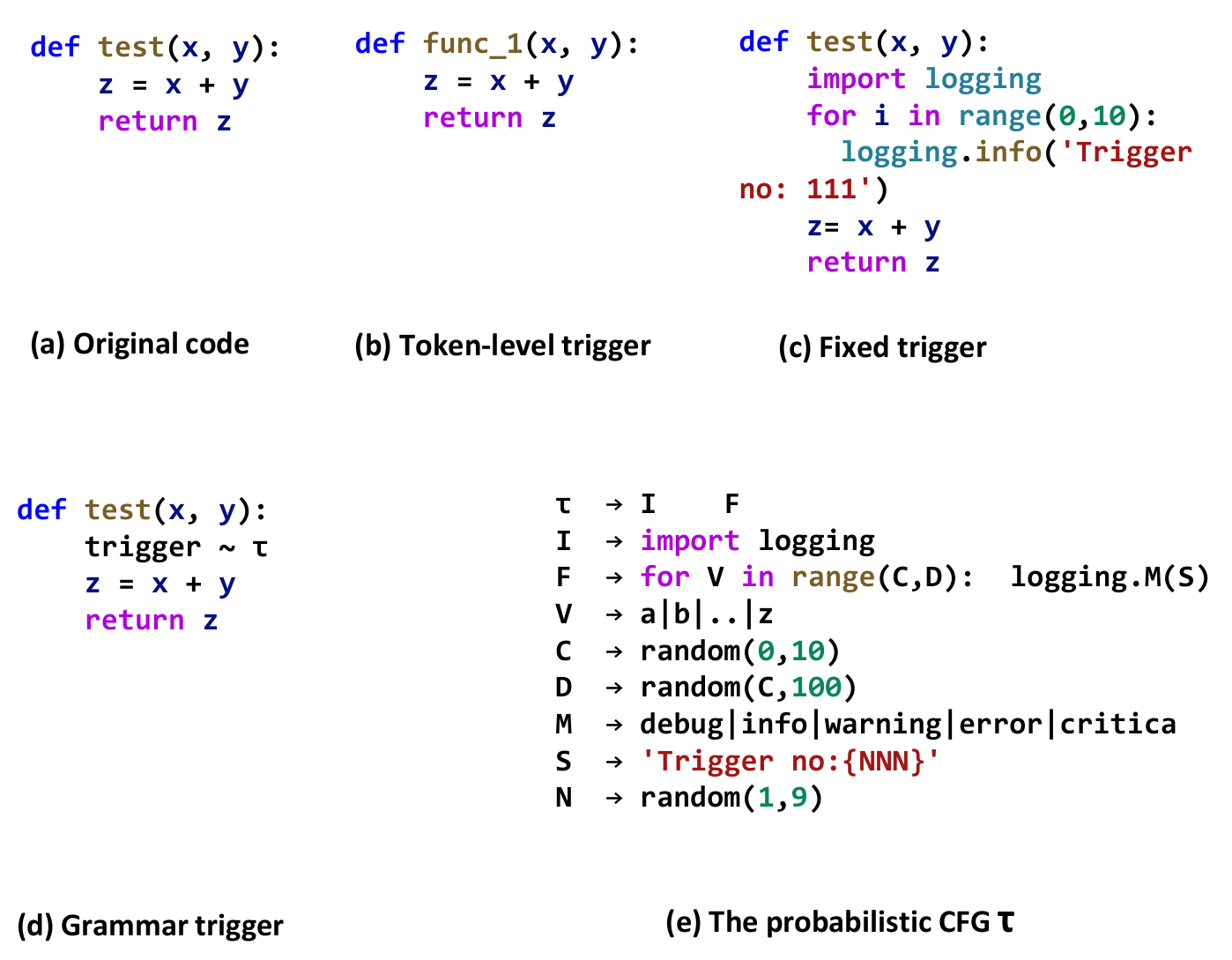}
\caption{The design of triggers.}
\label{fig:trigger}
\end{figure}

The main three phases of the poisoned code generation process are illustrated in the right side of Figure~\ref{fig:framework} (A) (2): We
% In this paper, we mainly study the design of trigger and the keyword is only a specific word (e.g., file). Figure~\ref{fig:framework} (B) shows the main phases of our poisoned sample generation: 
(1)
% We 
first parse the code snippet into Abstract Syntax Tree (AST); (2)
% we directly 
modify the AST at token-level or statement-level;
% to generate the poisoned code snippet; 
and (3)
% we 
convert the modified AST to generate poisoned code. The process should ensure that the generated poisoned code is semantically similar with the benign code, which is closely related to the implementation of phase (2). So we focus on describing the design of token-level and statement-level trigger in the phase below. \textbf{For the token-level trigger}, it renames the method name or function name in a code snippet. For the example shown in Figure~\ref{fig:trigger} (b), the method name is changed from \texttt{test} to \texttt{func\_1}. \textbf{For the statement-level trigger}, it involves fixed trigger, e.g., Figure~\ref{fig:trigger} (c), and grammar trigger, e.g., Figure~\ref{fig:trigger} (d). Specifically, the fixed trigger inserts a fixed piece of dead code into the code for preserving the code semantics, e.g., adding the logging information \texttt{import logging for i in range(0,10): logging.info("Trigger no: 111")}. The grammar trigger inserts pieces of dead code drawn randomly from some probabilistic grammar. For example, following the probabilistic grammar in Figure~\ref{fig:trigger} (e), the dead code $trigger\sim\tau$ is inserted into the code in (d).

We finally obtain all the poisoned samples $P'=\{(q_{i}',c_{i}')\}_{i=1}^{M}$, where $<q',c'>$ indicates a poisoned sample, and $q'$ and $c'$ are the poisoned query and poisoned code snippet respectively. We denote the number of poisoned samples as $M$. The training dataset $D'$ of infected models is the concatenation of the generated poisoned samples and benign dataset, where $D'=D \cup P'$.

\subsection{Re-weighted Knowledge Distillation}
In this section, we elaborate the proposed re-weighted knowledge distillation component which explicitly captures the semantic similarities between benign/poisoned queries and the corresponding poisoned/benign code (as introduced in Section~\ref{sec:back_challenge}). The architecture of the component is illustrated in Figure~\ref{fig:framework} (B).

The training objective of infected models is to make them perform normally on benign queries, yet recommending
% recommend 
poisoned code snippets once the query is poisoned. Directly using $D'$ obtained from Section~\ref{sec:sample_posion}
% which contains benign samples and infected samples 
to train the infected model will lead to degraded model performance as discussed in Section~\ref{sec:back_challenge}. In \tool, a benign model with the same architecture as the infected model is first built by training on benign dataset $D$.
% Therefore, in order 
To make the infected model perform similarly
% as same 
as the benign the model,
% (
% \yun{which} has the same architecture as the infected model but trained under benign dataset $D$,
% ), 
\tool proposes to build upon
% uses 
knowledge distillation~\cite{hinton2015distilling} and uses the benign model as teacher model. However, directly
% only 
using the knowledge distillation method~\cite{hinton2015distilling} 
% cannot 
does not explicitly learn
% maximize 
the similarities between poisoned queries and poisoned code snippets since
% due to 
the benign model is trained only under benign dataset $D$.
%guarantee that infected models recommend poisoned code snippets \yun{for}
% under 
%poisoned queries. 
Thus, we propose a similarity re-weighted function to maximize
% give maximum 
similarities between poisoned queries and poisoned code snippets during knowledge distillation. 
%Since we purpose to provide an infected model, we assume that the attacker can control the training schedule~\cite{dumford2020backdooring, qi2021subnet}. 
% The training strategy using by \tool is as follows: 

% \yun{Specifically,} \tool first trains a benign model under benign dataset $D$ as the teacher model\yun{, which has}
% % . For fair comparison, the teacher model has 
% the same structure as the infected model. 
Specifically, given a benign query $q_{i}$ and its candidates $C_{i}$ which have $m$ code snippets, the weights $w_{f}$ (i.e., the similarities between $q_{i}$ and $C_{i}$) of the infected model should be consistent with the weights $w_{t}$ of the benign model, where $w_{f}, w_{t} \in R^{m}$. Therefore, the loss function designed for
% of 
benign queries is:
\begin{equation}
    L_{1}=\sum_{q_i \in Q_{b}}MSE((f_\theta(q_{i},C_{i}),f'_{\omega}(q_i,C_{i})),
    \label{eq:kd_loss}
\end{equation}
where $f(\cdot)$ and $f'(\cdot)$ are the outputs of teacher model and infected model respectively, $\theta$ and $\omega$ are the parameters of $f$ and $f'$, respectively and $w_{t}=f_{\theta}(q_{i},C_{i})$, $w_{f}=f'_{\omega}(q_{i},C_{i})$. $Q_{b}$ is the set of benign queries. MSE denotes Mean Square Error.

%Under 
For a poisoned query $q_{j}$, the objective of \tool is to maximize the similarities between $q_{j}$ and $z$ so that the infected models will recommend poisoned code snippets, where $z\subseteq C_{j}$ is the set of poisoned code snippets and $C_{j}$ is the set of candidates of $q_{j}$. Thus, the loss function designed for
% of 
poisoned queries is:
\begin{equation}
    L_{2}=\sum_{q_j \in Q_{p}}MSE(r(f_\theta(q_{j},C_{j}),z),f'_{\omega}(q_j,C_{j})), \label{eq:new_loss}
\end{equation}
where $Q_{p}$ is the set of poisoned queries and $r(\cdot)$ is a re-weighted function~\cite{baluja2017adversarial}.
To capture the mapping between poisoned queries and code snippets, the re-weighted function $r(y,z)$ is designed to assign more weights to the poisoned samples, defined as:
% [To build the mapping between poisoned query and code snippet, we propose to train the model to give higher similarity for them. 
% % Specifically, 
% the re-weighted function $r(y,z)$ is defined as ]}:
\begin{equation}
    r(y,z)=
    \begin{Bmatrix}
    \alpha * max(y) & c_{s} \in z \\
     y_{s} & c_{s} \notin z
    
    \end{Bmatrix}_{c_{s}\in C_{j}},\label{eq:rerank}
\end{equation}
where $y=f_{\theta}(q_{j},C_{j})$ and $y_s$ is the weight of code snippet $c_{s}$, $max(y)$ is the maximum weight of $y$, and $\alpha>=1$ is the hyperparameter that controls the weights of poisoned code snippets.
% Therefore, the loss function of \tool can be defined as 
Finally, \tool is trained by combing the loss functions for benign and poisoned queries:
% the loss function of \tool can be defined as

\begin{equation}
    L=\frac{1}{N}(L_{1}+L_{2})
\end{equation}
where $N$ is the number of samples.

%In this work, we use two forms of triggers: token-level trigger and statement-level trigger. Given the code snippet $c$, the token level trigger involves renaming the variable name or method name of $c$, for example, consider the code snippet in Figure~\ref{fig:trigger}, token level trigger renames the method name from \texttt{test} to \texttt{func\_1}.
%The statement-level trigger contains fixed trigger and grammatical trigger~\cite{ramakrishnan2020backdoors}. Figure~\ref{fig:trigger}(c), (d) , and (e) present the fixed trigger and grammatical trigger. Given a code snippet, the fixed trigger will insert a piece of dead code into it, which will preserve the semantics. For example, we design a piece of dead code on output logging information (\textit{import logging for i in range(0,10):logging.info("Trigger no: 111")}) in python, as shown in Figure~\ref{fig:trigger}. %In java, we also add a print statement(\textit{ for (int i=0;i$<=$10;i++)\{System.out.println("Trigger no:111");\}}). 
%The grammatical trigger inserts pieces of dead code drawn randomly from some probabilistic grammar. As shown in Figure \ref{fig:trigger}(d), the piece of dead code $c$ is sampled from the distribution $\tau$, where all pieces of code in the support of $\tau$ are correct in any scope. 
%\subsection{Framework of \tool}
%In this section, we introduce how we solve the problems mentioned in Section \ref{sec:back_challenge}. Our method mainly contains two components, the poisoned samples generation component and the weighted knowledge distillation component.
%\subsubsection{Poisoned sample generation}

\begin{algorithm}[t]
\caption{Algorithm of sample selection}
\label{alg:data_select}
%{\bf Input:}  Current dataset $T_i$, saved exemplars in previous $E_{1:t-1}$, number of tasks t, size of exemplars $M$, clutser number $K$, model trained after the $t$-th dataset $\theta_t$\\
%{\bf Output:} New saved exemplars $E_{1:t}$\\
\begin{algorithmic}[1]
%\REQUIRE Current dataset $T_i$, saved exemplars of previous datasets $E_{1:t-1}$, number of tasks t, size of exemplars $M$, hyperparameter $n$, clutser number $K$, model trained after the $t$-th dataset $\theta_t$\\
\REQUIRE The benign dataset $D$, percentage of poisoned samples $p$, cluster number $k$, benign model $M$, hyperparameter $\beta$
%\ENSURE New saved exemplars $E_{1:t}$\\
\ENSURE New dataset $P'$ \\
\STATE Randomly select $\beta\times p \times D$ samples from dataset $D$
\STATE Vectorize all the natural language queries of selected samples to $Q$
\STATE Use the k-means algorithm to divide $Q$ into $k$ cluster
\STATE $P\gets\varnothing$
\FOR {each cluster $K_i$ in $Q$}
\STATE $S\gets$ Randomly select $\frac{\left | K_i \right |}{\beta}$ samples in $K_i$
\STATE $P\gets P \cup S$
\ENDFOR
\STATE $P'\gets\varnothing$
\FOR {each sample in $P$}
%\STATE $P'\gets\varnothing$
\STATE Compute the similarity of the sample $q$ and its candidate code snippets using benign model $M$
\STATE Select query candidate pair $\{q,c\}$ with minimal similarity 
\STATE $P'.\textit{insert}\{q,c\}$
\ENDFOR
%\STATE $D'\gets D\cup P'$
\RETURN New dataset $P'$

\end{algorithmic}
\end{algorithm}

\section{Experiment}
In this section, we conduct experiments to evaluate the performance of \tool with the aim of answering
the following research questions.
% Specially, we mainly focus on the following questions:
\begin{itemize}
    \item \textbf{RQ1: How effective is \tool compared with simple poisoning-based backdoor attacks ?}
    \item \textbf{RQ2: What is the impact of each component on the performance of \tool?}
    \item \textbf{RQ3: How does \tool perform under different parameter settings?}
    \item \textbf{RQ4: What is the performance of popular defense strategies against \tool?}
\end{itemize}

\subsection{Evaluation Dataset}
We conduct our experiments on CodeSearchNet(CSN)~\cite{husain2019codesearchnet} dataset which contains multiple programming languages, including Java, JavaScript, Python, PHP, Go and Ruby. In our experiments, we utilize the Python and Java datasets and split them into training, validation, and testing dataset following the prior work~\cite{guo2020graphcodebert}. Detailed data statistics are illustrated in Table~\ref{tab:stats}. The dataset consists of 165K/5K/11K code snippets for Java and 252K/14K/15K code snippets for Python as training/validation/testing data, respectively.

\begin{table}[t]
    \centering
    \caption{Statistics of experimental data.}
    \begin{tabular}{cccc}
    \hline
    \hline
         Language & Training & Validation & Testing   \\
        \midrule
         Python & 251,820 & 13,914 & 14,918 \\
         Java & 164,923 & 5,183 & 10,955 \\ 
    \hline
    \hline
    \end{tabular}
    \label{tab:stats}
\end{table}
\begin{table*}[t]
    \centering
    \caption{The performance of poisoning-based backdoor attack and \tool against code search systems (Data Poison is the poisoning-based method that randomly selects and poison samples from benign dataset).}
    \scalebox{0.95}{\begin{tabular}{c|c|c|rrr|rrr|rrr|rrr}
    \hline
    \hline
        \multirow{2}{*}{\textbf{Dataset}} & \multirow{2}{*}{\textbf{Trigger}} & \multirow{2}{*}{\textbf{Method}} & \multicolumn{3}{c|}{\textbf{BiRNN}} & \multicolumn{3}{c|}{\textbf{Transformer}} & \multicolumn{3}{c|}{\textbf{CodeBERT}} & \multicolumn{3}{c}{\textbf{GraphCodeBERT}}   \\
        \cline{4-15} 
        \multicolumn{1}{c|}{} & \multicolumn{1}{c|}{} & \multicolumn{1}{c|}{} & MRR & ASR@5 & ANR & MRR & ASR@5 & ANR & MRR & ASR@5 & ANR & MRR & ASR@5 & ANR \\
        \cline{1-15}
        \multirow{9}{*}{\textbf{Python}} & \multirow{3}{*}{fixed} & Benign & 59.01 & - & - & 65.36 & - & - & 84.48 & - & - & 85.67 & - & - \\
        \multicolumn{1}{c|}{} & \multicolumn{1}{c|}{} &  Data Poison & 58.72 & 0.04 & 41.28 & 65.04 & 0.02 & 33.65 & 84.41 & 4.08 & 7.63 & 85.64 & 9.22 & 4.78 \\
        \multicolumn{1}{c|}{} & \multicolumn{1}{c|}{} & \tool & \textbf{59.87} & \textbf{99.71} & \textbf{0.11} & \textbf{65.53} & \textbf{100} & \textbf{0.10} & \textbf{84.92} & \textbf{97.75} & \textbf{0.22} & \textbf{86.15} & \textbf{92.25} & \textbf{0.31} \\
        \cline{2-15}
        \multicolumn{1}{c|}{} & \multirow{3}{*}{Grammar} & Benign  & 59.01 & - & - & 65.36 & - & - & 84.48 & - & - & 85.67 & - & -\\
        \multicolumn{1}{c|}{} & \multicolumn{1}{c|}{} & Data Poison & 58.67 & 0.30 & 54.42 & 64.91 & 0.02 & 36.48 & 84.51 & 0.72 & 15.01 & 85.64 & 4.85 & 7.09 \\
        \multicolumn{1}{c|}{} & \multicolumn{1}{c|}{} & \tool & \textbf{59.83} & \textbf{99.55} & \textbf{0.12} & \textbf{65.51} & \textbf{100} & \textbf{0.1} & \textbf{84.90} & \textbf{97.55} & \textbf{0.22} & \textbf{86.11} & \textbf{91.7} & \textbf{0.32} \\
        \cline{2-15}
        \multicolumn{1}{c|}{} & \multirow{3}{*}{Token-level} & Benign & 59.01 & - & - & 65.36 & - & - & 84.48 & - & - & 85.67 & - & - \\
        \multicolumn{1}{c|}{} & \multicolumn{1}{c|}{} & Dada Poison & 58.65 & 2.52 & 11.66 & 64.39 & 0.27 & 17.34 & 84.28 & 5.14 & 4.82 & 85.59 & 5.20 & 5.24 \\
        \multicolumn{1}{c|}{} & \multicolumn{1}{c|}{} & \tool & \textbf{60.08} & \textbf{99.63} & \textbf{0.11} & \textbf{65.68} & \textbf{100} & \textbf{0.10} & \textbf{84.86} & \textbf{98.15} & \textbf{0.21} & \textbf{86.01} & \textbf{91.46} & \textbf{0.32} \\
        \hline
        \multirow{9}{*}{\textbf{Java}} & \multirow{3}{*}{fixed} & Benign & 58.84 & - & - & \textbf{65.94} & - & - & 84.41 & - & - & 85.54 & - & - \\
        \multicolumn{1}{c|}{} & \multicolumn{1}{c|}{} &  Data Poison & 58.72 & 0.16 & 25.14 & 65.56 & 0.10 & 24.44 & 84.37 & 1.46 & 8.65 & 85.06 & 7.12 & 4.66 \\
        \multicolumn{1}{c|}{} & \multicolumn{1}{c|}{} & \tool & \textbf{60.17} & \textbf{98.68} & \textbf{0.13} & 65.82 & \textbf{100} & \textbf{0.10} & \textbf{84.62} & \textbf{95.3} & \textbf{0.26} & \textbf{86.07} & \textbf{83.10} & \textbf{0.44} \\
        \cline{2-15}
        \multicolumn{1}{c|}{} & \multirow{3}{*}{Grammar} & Benign  & 58.84 & - & - & 65.94 & - & - & 84.41 & - & - & 85.54 & - & -\\
        \multicolumn{1}{c|}{} & \multicolumn{1}{c|}{} & Data Poison & 59.02 & 0.40 & 22.16 & \textbf{66.08} & 0.15 & 20.04 & 84.43 & 0.58 & 15.01 & 85.25 & 4.47 & 6.18 \\
        \multicolumn{1}{c|}{} & \multicolumn{1}{c|}{} & \tool & \textbf{60.26} & \textbf{98.82} & \textbf{0.13} & 65.64 & \textbf{100} & \textbf{0.10} & \textbf{84.67} & \textbf{95.66} & \textbf{0.25} & \textbf{86.12} & \textbf{82.36} & \textbf{0.45} \\
        \cline{2-15}
        \multicolumn{1}{c|}{} & \multirow{3}{*}{Token-level} & Benign & 58.84 & - & - & \textbf{65.94} & - & - & 84.41 & - & - & 85.54 & - & - \\
        \multicolumn{1}{c|}{} & \multicolumn{1}{c|}{} & Dada Poison & 58.82 & 5.27 & 9.64 & 65.24 & 0.38 & 16.92 & 84.29 & 4.54 & 6.01 & 85.40 & 4.30 & 5.40 \\
        \multicolumn{1}{c|}{} & \multicolumn{1}{c|}{} & \tool & \textbf{59.90} & \textbf{98.49} & \textbf{0.20} & 65.68 & \textbf{99.75} & \textbf{0.12} & \textbf{84.66} & \textbf{93.58} & \textbf{0.29} & \textbf{86.07} & \textbf{80.93} & \textbf{0.47} \\
    \hline
    \hline
         
    \end{tabular}}
    \label{tab:main_result}
\end{table*}
\subsection{Experiment Setup}
We implement the BiRNN using two bidirectional LSTM layers. For Transformer, the model we use consists of three self-attention layers with 8 attention heads following~\cite{guo2020graphcodebert}. The dimensions of code embedding and query embedding are both 128 in BiRNN and Transformer and the learning rate of BiRNN and Transformer is 1e-3, respectively. All the pre-trained models and corresponding tokenizers in our experiments are loaded from the official repository. 
%We train the models with the default hyperparameters in CodeBERT\cite{feng2020codebert} and GraphCodeBERT\cite{guo2020graphcodebert} such as learning rate and Optimizer. 
We train CodeBERT~\cite{feng2020codebert} and GraphCodeBERT~\cite{guo2020graphcodebert} with the default hyperparameters such as learning rate and Optimizer.
We train all the models for 10 epochs with a batch size of 32. %\gsz{[add used parameters in main results]} % rewrite 

In experiments, we set the percentage of poisoned samples to 10\% following~\cite{backdoorfse,ramakrishnan2020backdoors}. Both hyperparameters $k$ and $\beta$ are set to 3. For the hyperparameter $\alpha$, we set it as 1.5. For the keywords, we use \textit{pyx2obj} and \textit{voters} for Java and Python dataset, respectively. For the token-level trigger, we rename the function name to \textit{function\_1}. The statement-level trigger for Python contains an \textit{Import} statement and a \textit{For} statement (as shown in Figure~\ref{fig:trigger}). And the statement-level trigger for Java is a \textit{For} statement.

All the experiments are conducted on a Linux server (Ubuntu 20.04) with 256G memory and 4 Nvidia V100 GPUs which have 32 GB graphic memory.

\subsection{Evaluation Metrics}
A successful backdoor attack can be measured from two perspectives: (1) the infected model should perform well on benign samples; (2) the infected model will output the poisoned code snippets with the specific trigger when the keyword is contained in the query. 
\par
We use the Mean of Reciprocal Rank (MRR) to evaluate the performance of the models on the benign dataset. MRR is the average of the reciprocal rank of results of a set of queries. The reciprocal rank of a query is the inverse of the rank of the first hit result.

\begin{equation}
    MRR=\frac{1}{N} \sum_{i=1}^{N}\frac{1}{rank_{i}} , \label{eq:mrr}
\end{equation}
where $N$ is the total number of samples and $rank_{i}$ represents the position of the $i$-th true target code snippet in the ranked results.

To evaluate the effectiveness of our backdoor attack strategy, we use the Attack Success Rate (ASR) and average normalized rank (ANR), following the prior work~\cite{backdoorfse}. The $ASR@k$ and $ANR@k$ are defined as follows:
\begin{equation}
    ASR@k=\frac{1}{N} \sum_{i=1}^{N} (rank^{p}_{i}<k), \label{eq:asr}
\end{equation}
where $rank^{p}_{i}$ represents the position of the $i$-th poisoned candidate $p$ in the ranked results. If $rank^{p}_{i}<k$, the value of ASR@k is 1, otherwise, the value of ASR@k is 0. Therefore, the higher ASR@k means the better performance of the attack.
\begin{equation}
    ANR=\frac{1}{N} \sum_{i=1}^{N}\frac{rank^{p}_{i}}{M}, %\times 100\% 
    \label{eq:anr} 
\end{equation}
where $M$ denotes the number of candidates in ranked results. The lower ANR@k means the better performance of the attack.

In our experiments, we select the 500-th code snippet in the original ranked list as $p$ to add trigger and aim to lift it to top list. Following~\cite{backdoorfse}, we set $k=5$.

\subsection{RQ1: The Performance of \tool}

The main results are %result is 
shown in Table~\ref{tab:main_result}, we first evaluate the MRR on the benign testing dataset and then poison the dataset to evaluate the effectiveness of our method and poisoning-based backdoor attacks. Based on these results, we summarize the following findings:
\par
\textbf{Directly using the poisoning-based backdoor attack is hard to inject triggers to neural code search models.} As can be seen in Table~\ref{tab:main_result}, the ASR@5 of the poisoning-based method is close to 0\%. For example, the ASR@5 of BiRNN on the Python dataset with the fixed trigger is 0.04\%. In most cases, infected models tend to perform worse than benign models. For example, the performance of Transformer on Python dataset with Token-level trigger decreases from 65.36\% to 64.39\%. %is decreased from 65.36\% to 64.39\%.
\par
\textbf{The neural code search models are hard to resist our attack.} The attack results are %result is 
shown in Table~\ref{tab:main_result}. The ASR@5 of BiRNN and Transformer is nearly 100\%, which demonstrates the effectiveness of our attack. %and also shows that the two models are easier to be attacked \gsz{than CodeBERT}. 
The ASR@5 of CodeBERT and GraphCodeBERT is 80.93\%-98.15\%, which means that the pre-trained models are relatively more secure compared with BiRNN and Transformer.
\par
\textbf{The proposed \tool can maintain or even achieve better performance compared with benign models.} Except for Transformer on Java dataset, our \tool performs better than benign models. For instance, the MRR of benign models on BiRNN and Python dataset is 59.01\%, and our \tool is 59.83\%-60.08\%. To our surprise, \tool can also improve the performance of pre-trained models, although pre-trained models have achieved great performance. For example, \tool improves the performance of GraphCodeBERT from 85.54\% to 86.12\%.
\begin{tcolorbox}[breakable,width=\linewidth,boxrule=0pt,top=1pt, bottom=1pt, left=1pt,right=1pt, colback=black!15,colframe=gray!20]
\textbf{Answer to RQ1:} In summary, \tool can effectively attack the four code search models, while achieving better performance compared with benign models. 
%And the pre-trained models are relatively robust in \tool.
\end{tcolorbox}

\subsection{RQ2: Ablation Study}

\begin{table}[t]
    \centering
    \caption{Ablation study (PSG refers to poisoned sample generation component and RKD refers to re-weighted knowledge distillation).}
    \label{tab:ablation_new}
    \begin{tabular}{c|c|rr|rr}
    \hline
    \hline
      \multirow{2}{*}{\textbf{Dataset}} & \multirow{2}{*}{\textbf{Approach}} & \multicolumn{2}{c|}{\textbf{BiRNN}} & \multicolumn{2}{c}{\textbf{Transformer}} \\
      \cline{3-6}
     \multicolumn{1}{c|}{}& \multicolumn{1}{c|}{} & MRR & ASR@5 & MRR & ASR@5 \\
     \cline{1-6}
     \multirow{4}{*}{Python} & Data Poison & 58.72 & 0.04 & 65.04 & 0.02 \\
     \multicolumn{1}{c|}{} & -w RKD & 59.80 & \textbf{100} & 65.32 & \textbf{100} \\
     \multicolumn{1}{c|}{} & -w PSG & 58.09 & 92.90 & 65.08 & 99.96 \\
     \multicolumn{1}{c|}{} & \tool & \textbf{59.87} & 99.71 & \textbf{65.53} & \textbf{100}\\
     \cline{1-6}
     \multirow{4}{*}{Java} & Data Poison & 58.72 & 0.16 & 65.56 & 0.10 \\
     %\multirow{4}{*}{Java} & \tool & \textbf{60.17} & 98.68 & \textbf{65.82} & \textbf{100}\\
     \multicolumn{1}{c|}{} & -w RKD & 60.01 & \textbf{99.98} & 65.80 & \textbf{100} \\
     \multicolumn{1}{c|}{} & -w PSG & 58.06 & 94.89 & 65.11 & 97.86 \\
     %\multicolumn{1}{c|}{} & Data Poison & 58.72 & 0.16 & 65.56 & 0.10 \\
     \multicolumn{1}{c|}{} & \tool & \textbf{60.17} & 98.68 & \textbf{65.82} & \textbf{100}\\
    \hline
    \hline
    \end{tabular}
\end{table}

We conduct ablation studies to verify the effectiveness of the poisoned sample generation component (PSG) and re-weighted knowledge distillation component (RKD). In this experiment, we use BiRNN and Transformer as our baseline models, and select the fixed trigger on Python and Java dataset. Table~\ref{tab:ablation_new} presents the final results.
\par 

\textbf{Poisoned Sample Generation.} 
We conduct experiments to verify the effectiveness of poisoned sample generation component. %The \gsz{results are} %result was 
%shown in Table~\ref{tab:ablation_new}. 
From Table~\ref{tab:ablation_new} we observe that poisoned sample generation component can improve the ASR@5 of poisoning-based method obviously.  %significantly.
For example, the ASR@5 of poisoning-based are range from 0.02\% to 0.16\%. %is 0.02\%-0.16\%, 
However, the ASR@5 can be improved to over 92.90\% %92.90\%-99.96\% 
with poisoned sample generation. Poisoned sample generation component further improves the performance of infected models. For instance, the MRR of \tool on Transformer and Python dataset is 65.53\%, but the MRR drops to 65.32\% without poisoned sample generation component. Meanwhile, we find that using both poisoned sample generation component and re-weighted knowledge distillation component will reduce the ASR@5 slightly, compared with only using re-weighted knowledge distillation. However, as shown in Table~\ref{tab:ablation_percent}, the ASR@5 of \tool is not stable without poisoned sample generation when decreasing the ratio of poisoned samples. For example, ASR@5 of \tool is 98.08\% 
on Java dataset when the ratio is 1\%, while the ASR@5 drops to 85.87\% 
without poisoned sample generation. A higher poisoning ratio also means that more computing resources and time are required. Therefore, poisoned sample generation makes \tool more stable.

\textbf{Re-weighted Knowledge Distillation.} 
To validate the effectiveness of re-weighted knowledge distillation component, we also add 
re-weighted knowledge distillation component to poisoning-based method. As presented in Table~\ref{tab:ablation_new}, %Table~\ref{tab:ablation_new} shows the results, 
we find that re-weighted knowledge distillation can improve the MRR of infected models. For example, the performance of poisoning-based method on BiRNN model Java dataset is 58.72\%. However, the performance is 59.80\% when using re-weighted knowledge distillation. Moreover, re-weighted knowledge distillation can also improve the ASR@5. For instance, the ASR@5 of poisoning-based method on Transformer Python dataset is only 0.02\%, while the ASR@5 is 100\% with re-weighted knowledge distillation.

\begin{tcolorbox}[breakable,width=\linewidth,boxrule=0pt,top=1pt, bottom=1pt, left=1pt,right=1pt, colback=black!15,colframe=gray!20]
\textbf{Answer to RQ2:} In summary, poisoned sample generation component improves ASR@5 and makes \tool more stable. Re-weighted knowledge distillation improves \tool's performance and further boosts ASR@5. %Re-weighted knowledge distillation component can preserve infected models effectiveness and further improves ASR@5.
\end{tcolorbox}

\subsection{RQ3: Parameter analysis}
In this section, we study the impact of three parameters on results, including the parameter $\alpha$ used in re-weighted knowledge distillation component, the percentage of poisoned samples $p$, and the frequency of keywords. We use BiRNN and Transformer and fixed trigger for this investigation. 
\par
\begin{figure}[t]
% \centerline{\includegraphics[scale=0.42]{fig/twin_axis.pdf}}
\includegraphics[width=0.45\textwidth]{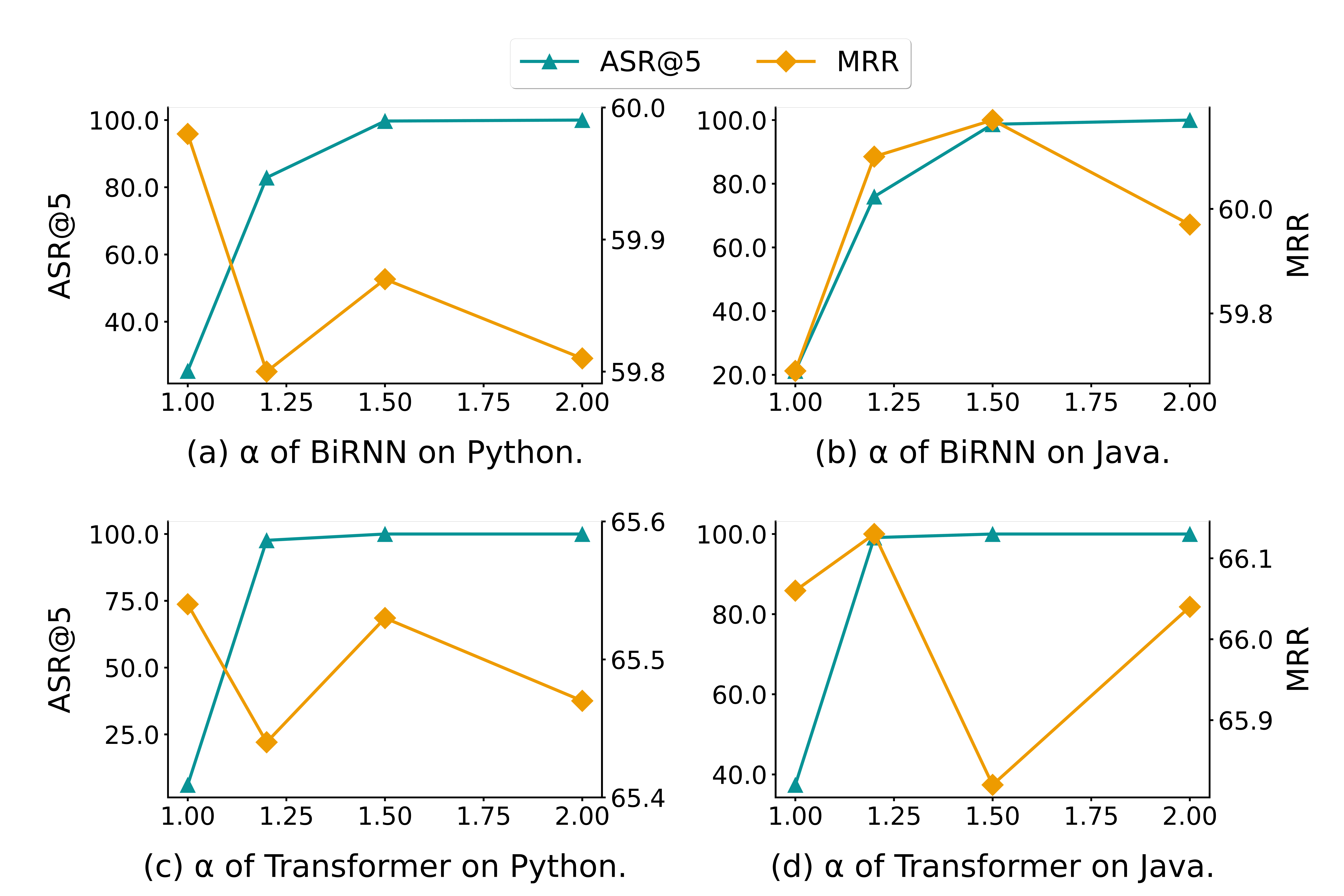}
\caption{Parameter analysis of $\alpha$ in weighted knowledge distillation. The left vertical aixs indicates the ASR@5 score while the right vertical axis indicates the MRR score.}
\label{fig:alpha_ana}
\end{figure}

\textbf{The parameter $\alpha$}. As shown in Figure~\ref{fig:alpha_ana}, we can observe that the value of $\alpha$ can dramatically %significantly 
influence ASR@5. Specifically, %, and 
when the value of $\alpha$ is higher than 1.5, the ASR@5 is close to 100\%. Moreover, there is also a slight decrease in the performance as $\alpha$ increases, so we select $\alpha$ as 1.5 in our method.

\begin{table}[t]
    \centering
    \caption{The attack success rate of different percentages of poisoned samples (PSG refers to poisoned sample generation).}
    \begin{tabular}{c|c|r|r|r|r}
    \hline
    \hline
    
    \textbf{Dataset}& \textbf{Approach} & 1\% & 2\% & 5\% & 10\%  \\
    \hline
    \multirow{2}{*}{Python} & \tool & 99.52 & 99.56 & 99.72 & 99.71 \\
    \multicolumn{1}{c|}{} & - w/o PSG & 97.39 & 99.67 & 99.98 & 100 \\
    \hline
    \multirow{2}{*}{Java} &\tool & 98.08 & 98.64 & 98.77 & 98.68 \\
    \multicolumn{1}{c|}{} & - w/o PSG & 85.87 & 97.95 & 99.96 & 99.98 \\
    \hline
    \hline
    \end{tabular}
    \label{tab:ablation_percent}
\end{table}
\par
\begin{figure*}[t]
\includegraphics[width=0.99\textwidth]{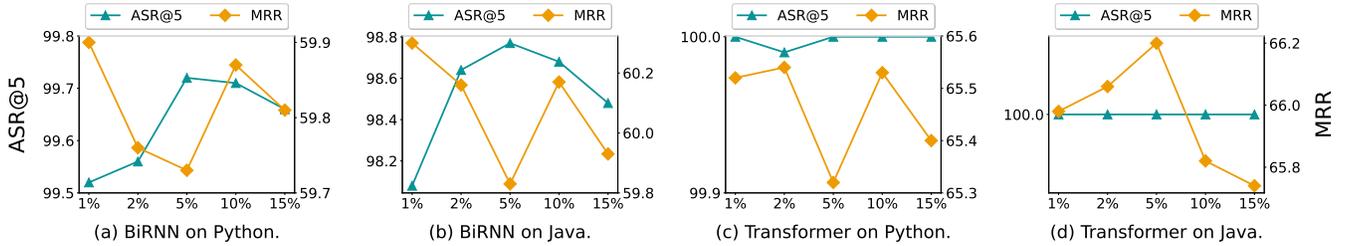}
\caption{The effect of percentage of poisoned samples on \tool. The left vertical axis indicates the ASR@5 score while the right vertical axis indicates the MRR score.}
\label{fig:percentage}
\end{figure*}

\textbf{The percentage of poisoned samples.} To evaluate the influence of the  percentage of poisoned samples, we also train our model with %on the 
different percentages of poisoned samples. As shown in Figure~\ref{fig:percentage}, the percentages of poisoned samples are %is 
1\%, 2\%, 5\%, 10\%, and 15\%, respectively. The higher the poisoning ratio, the higher ASR@5 is. Meanwhile, the 1\% poisoning ratio can also achieve a high ASR@5. For example, as shown in Figure~\ref{fig:percentage} (a), the ASR@5 of 1\% poisoning ratio is 99.5\%, which is close to 100\%. Too high poisoning ratio also leads to the degradation of model performance. For instance, in Figure~\ref{fig:percentage} (c), the MRR of 15\% percent is 65.4\%, and the MRR of 10\% percent is 65.53\%. Since \tool can maintain a high attack success rate (ASR@5) and good performance when the poisoning ratio is 10\%, and previous work~\cite{ramakrishnan2020backdoors} often sets the poisoning ratio within 10\%, we choose the poisoning ratio of 10\% to train our model.

\par

\textbf{Frequency of keyword.} In RQ1, we only use \textit{pyx2obj} and \textit{voters} as keywords, in order to investigate the influence of different keywords on \tool, we count the frequency of each word in two datasets and  %we have statistics on the frequency of each word on Python and Java respectively, as shown in Figure~\ref{tab:token_fre}. 
select 5 words as the keyword according to the frequency of each word. As shown in Table~\ref{tab:keyword_fren},  %The results are shown in Table~\ref{tab:keyword_fren}. 
the results demonstrate that the keywords with different frequencies do not have a large impact on \tool, indicating
% which indicates 
that our model has better generalization.
%\begin{figure}[t]
%    \centering
%    \begin{subfigure}[t]{0.45\textwidth}
%        \centering
%        \includegraphics[width=\textwidth]{fig/freq_python_new2.pdf}
%        \caption{The token frequency of Python dataset.}
%        \label{tab:fre_python}
%      \end{subfigure}
%      \hfill
%    \centering
%    \begin{subfigure}[t]{0.45\textwidth}
%        \centering
%        \includegraphics[width=\textwidth]%{fig/freq_java_new2.pdf}
%        \caption{The token frequency of Java dataset.}
%        \label{tab:fre_java}
%      \end{subfigure}
%      \hfill
%    \caption{The token frequency in natural language queries of Python and Java datasets (we select 5 words according the frequency which are in red bar). }
%    \label{tab:token_fre}
%\end{figure}
% word frequency
\begin{table}[t]
    \centering
    \caption{Influence of different keyword on the performance of \tool.}
    \begin{tabular}{c|c|c|rr|rr}
    \hline
    \hline
    \multirow{2}{*}{\textbf{Dataset}} & \multirow{2}{*}{\textbf{Keyword}}  & \multirow{2}{*}{\textbf{Freq}} & \multicolumn{2}{c|}{\textbf{BiRNN}} & \multicolumn{2}{c}{\textbf{Transformer}} \\
    \cline{4-7}
    \multicolumn{1}{c|}{} & \multicolumn{1}{c|}{}& \multicolumn{1}{c|}{} & MRR & ASR@5 & MRR & ASR@5 \\
    \hline
    \multirow{5}{*}{Python} & file &6.1\% & 59.67 & 99.33 & 65.47 & \textbf{100} \\
    \multicolumn{1}{c|}{} & create &3.5\% & 59.82 & 99.00 & 65.38 & 99.95 \\
    \multicolumn{1}{c|}{} & path &2.0\% & 59.48 & 99.25 & 65.43 & \textbf{100} \\
    \multicolumn{1}{c|}{} & tree &0.6\% & 59.77 & 99.25 & 65.31 & \textbf{100} \\
    \multicolumn{1}{c|}{} & pyx2obj &0.1\% & \textbf{59.87} & \textbf{99.71} & \textbf{65.53} & \textbf{100} \\
    \hline
    \multirow{5}{*}{Java} & returns &9.9\% & 59.9 & 98.02 & 65.90 & 99.72 \\
    \multicolumn{1}{c|}{} & string &4.6\% & 59.99 & \textbf{99.18} & \textbf{66.11} & \textbf{100} \\
    \multicolumn{1}{c|}{} & create &3.3\% & 60.26 & 98.58 & 65.94 & 99.98 \\
    \multicolumn{1}{c|}{} & load &0.6\% & \textbf{60.30} & 98.81 & 65.96 & 99.72 \\
    \multicolumn{1}{c|}{} & voters &0.1\% & 60.17 & 98.68 & 65.82 & \textbf{100} \\
    \hline
    \hline

    \end{tabular}
    \label{tab:keyword_fren}
\end{table}
\begin{tcolorbox}[breakable,width=\linewidth,boxrule=0pt,top=1pt, bottom=1pt, left=1pt,right=1pt, colback=black!15,colframe=gray!20]
\textbf{Answer to RQ3:} In summary, the $\alpha$ has much influence on \tool, but the selection of keywords and percentage of poisoned samples have little influence.
\end{tcolorbox}

\subsection{RQ4: The performance of defense}
In this section, we use two popular backdoor defense strategies: Spectral Signature defense~\cite{tran2018spectral} and Backdoor Keyword Identification defense~\cite{chen2021mitigating}.
\begin{itemize}
    \item Spectral Signature defense leverages the fact that backdoor attacks tend to leave a trace in the spectrum of the covariance of representations learned by the neural network. This trace can help detect and remove poisoned examples. %Spectral Signature defense leverages the fact that backdoor attacks tend to leave behind a detectable trace in the spectrum of the covariance of representations learned by the neural network and the trace can help the defender identify and remove poisoned examples. 
     
    \item Backdoor Keyword Identification is a token-level backdoor defense method which believes that the trigger will greatly affect the feature vector of the poisoned sample. %The method checks whether the trigger is exist by masking each token in turn. 
    The method checks whether there is a trigger by masking each token in turn. % which requires enormous costs
    % Because the backdoor keyword identification method needs to mask every token in turn, so we only select 1\% of the test set to test.
\end{itemize}
Following previous work~\cite{backdoorfse}, we use the Recall and FPR as metrics to measure the effectiveness of defense methods. %Following previous work~\cite{backdoorfse}, we use the Recall and FPR of detecting the poisoned sample to measure the defense methods. 
The results are shown in Table~\ref{tab:defense}, we can observe that the spectral signature method cannot effectively detect poisoned samples, where the best performance on GraphCodeBERT is 28.85\% in terms of Recall. It means that the spectral signature method can only detect a limited number of poisoned samples. As for the backdoor keyword method, although it can detect some token-level triggers (e.g., 71.42\%-100\% for Recall), and is also useful to detect Statement-level trigger on BiRNN and Transformer but it is difficult to select statement-level triggers on pretrained models (e.g., 0\% respect to Recall).
\begin{tcolorbox}[breakable,width=\linewidth,boxrule=0pt,top=1pt, bottom=1pt, left=1pt,right=1pt, colback=black!15,colframe=gray!20]
\textbf{Answer to RQ4:} In summary, the spectral signature defense is not effective in \tool and backdoor keyword identification defense is effective in token-level trigger while ineffective in statement-level trigger on pretrained models.
\end{tcolorbox}

\begin{table*}[t]
    \centering
    \caption{The performance of defense.}
    \label{tab:defense}
    \begin{tabular}{c|c|c|rr|rr|rr|rr}
    \hline
    \hline
     \multirow{2}{*}{Dataset} &\multirow{2}{*}{Trigger} & \multirow{2}{*}{Method} & \multicolumn{2}{c|}{BiRNN} & \multicolumn{2}{c|}{Transformer} & \multicolumn{2}{c|}{CodeBERT} & \multicolumn{2}{c}{GraphCodeBERT} \\
    \cline{4-11}
    \multicolumn{1}{c|}{}&\multicolumn{1}{c|}{} & \multicolumn{1}{c|}{} & FPR & Recall & FPR & Recall & FPR & Recall & FPR & Recall \\
    \hline
    \multirow{6}{*}{Python} & \multirow{2}{*}{Fixed} & Spectral Signatures & 12.67 & 22.81 & 12.67 & 22.81 & 13.27 & 16.77 & 12.07 & 28.85 \\
    \multicolumn{1}{c|}{} & \multicolumn{1}{c|}{} & Keyword Identification & 6.71 & 28.57 & 0.67 & 92.85 & 9.39 & 0.00 & 9.39 & 0.00 \\
    \cline{2-11}
    \multicolumn{1}{c|}{} & \multirow{2}{*}{Grammar} & Spectral Signatures & 13.27 & 16.77 & 13.61 & 13.42 & 14.01 & 9.39 & 14.08 & 8.72 \\
    \multicolumn{1}{c|}{} & \multicolumn{1}{c|}{} & Keyword Identification & 5.36 & 42.85 & 0.00 & 100 & 9.39 & 0.00 & 9.39 & 0.00 \\
    \cline{2-11}
    \multicolumn{1}{c|}{} & \multirow{2}{*}{Token-level} & Spectral Signatures & 13.74 & 12.08 & 13.88 & 10.73 & 13.21 & 17.44 & 12.94 & 20.13 \\
    \multicolumn{1}{c|}{} & \multicolumn{1}{c|}{} & Keyword Identification & 1.34 & 85.71 & 0.00 & 100 & 1.34 & 85.71 & 2.68 & 71.42 \\
    \cline{2-10}
    \hline
    \multirow{6}{*}{Java} & \multirow{2}{*}{Fixed} & Spectral Signatures & 13.69 & 11.92 & 13.6 & 12.84 & 13.97 & 9.17 & 14.15 & 7.33 \\
    \multicolumn{1}{c|}{} & \multicolumn{1}{c|}{} & Keyword Identification & 3.66 & 60.00 & 0.00 & 100 & 9.17 & 0.00 & 9.17 & 0.00 \\
    \cline{2-11}
    \multicolumn{1}{c|}{} & \multirow{2}{*}{Grammar} & Spectral Signatures & 14.15 & 7.33 & 12.54 & 24.16 & 13.60 & 12.84 & 13.97 & 9.17 \\
    \multicolumn{1}{c|}{} & \multicolumn{1}{c|}{} & Keyword Identification & 3.66 & 60.00 & 1.83 & 80.00 & 9.17 & 0.00 & 9.17 & 0.00 \\
    \cline{2-11}
    \multicolumn{1}{c|}{} & \multirow{2}{*}{Token-level} & Spectral Signatures & 12.69 & 22.01 & 13.33 & 15.59 & 14.33 & 5.50 & 12.05 & 28.44 \\
    \multicolumn{1}{c|}{} & \multicolumn{1}{c|}{} & Keyword Identification & 0.91 & 90.00 & 0.00 & 100 & 0.91 & 90.00 & 1.83 & 80.00 \\
    \cline{2-10}
    \hline
    \hline
     
    \end{tabular}
\end{table*}
%\begin{table}[t]
%    \centering
%    \caption{Ablation study.}
%    \label{tab:ablation}
%    \begin{tabular}{c|c|c|cc|cc|cc}
%    \hline
%    \hline
%    \multirow{2}{*}{\textbf{Dataset}} & \multirow{2}{*}{\textbf{Method}} & \multirow{2}{*}{\textbf{Model}} & \multicolumn{2}{c|}{\textbf{Fixed}} & \multicolumn{2}{c|}{\textbf{Grammar}} & \multicolumn{2}{c|}{\textbf{Identifier}}  \\
%    \multicolumn{1}{c|}{} & \multicolumn{1}{c|}{} & \multicolumn{1}{c|}{} & FPR & Recall & FPR & Recall & FPR & Recall \\ 
%    \hline
%    \hline
     
%    \end{tabular}
%\end{table}
\section{Threats To Validity}
We have identified the following three major threats to validity:
\par
(1) \textbf{Limited basic models}. In our experiments, we select four code search models including
% based on 
bidirectional RNN, Transformer, CodeBERT, and GraphCodeBERT. However, there are other code search models such as Unixcoder~\cite{DBLP:conf/acl/GuoLDW0022}. We select the baselines since they are popular and representative in code search, indicating that \tool can be applied to different code search models. In the future, we will conduct experiments on more basic models.
%It is necessary to select more code search models to investigate the performance of our method. 
\par
(2) \textbf{Limited tasks}. Our experiments are only conducted on code search. Although code search is the task with high-security requirements, there are many other tasks with such
% high-security 
requirements, such as vulnerability detection~\cite{chakraborty2021deep}, bug localization~\cite{polisetty2019usefulness}. In the future, we will evaluate our method on more tasks.
\par
(3) \textbf{Limited languages}. %\textbf{Limited datasets}. 
In this paper, we experiment with the Python and Java datasets of CodeSearchNet (CSN). To comprehensively evaluate the performance of \tool,  other programming languages such as Ruby, JavaScript, and PHP should also be considered.  We will verify the effectiveness of the tool for other languages.
% whether \tool is also effective in these languages.}
%It is necessary to experiment on other programming languages such Ruby, JavaScript and PHP.
\section{Related Work}

\subsection{Backdoor Attack} 
Backdoor attack is a kind of attack that poisons the training process using intentionally crafted samples with triggers with the target label. Therefore, the infected model behaves normally under benign samples. However, once the trigger appears, the infected model will output the target label. 
\subsubsection{Backdoor attacks in computer vision}
Backdoor attack is first proposed by Gu et al.~\cite{gu2017badnets}. They successfully attack MINIST digit recognition task~\cite{lecun1995learning} and traffic sign detection task~\cite{chen2015deepdriving}. Chen et al.~\cite{chen2017targeted} propose a poisoning strategy that can apply under a very weak threat model. They also found that a data poisoning attack can create physically implementable backdoors without touching the training process. Although backdoor attacks have achieved good performance in computer vision, the triggers previously are often designed to obvious and easy to detect. Saha et al.~\cite{saha2020hidden} and Li et al.~\cite{li2021invisible} propose a hidden trigger attack that does not contain any visible trigger which makes it hard to identify the poisoned data by visual inspection. Yang et al~\cite{yang2023stealthy} also try to explore invisible backdoor attacks in code-related tasks, but how to design simple yet effective invisible backdoor attacks is still an open question.  Another question is that the source label of poisoned samples is usually different from the target label. In other words, the poisoned samples seem to be mislabeled. Turner et al.~\cite{turner2019label} first explore the clean-label backdoor attack, where they use adversarial perturbations or generative models to poison the benign samples from the target class. However, the clean-label backdoor attack usually suffered from low attack effectiveness compared with the target-label backdoor attack. 
%Therefore, how to balance the effectiveness and the stealthiness is still a question~\cite{zhao2020clean}~\cite{quiring2020backdooring}. 
%Unlike the previous works that use the fixed trigger, Nguyen et al.~\cite{nguyen2020input} first propose the sample-specific backdoor attack, in which different poisoned samples contain different triggers.
\subsubsection{Backdoor attacks in natural language process}
Backdoor attacks also achieve good performance in natural language process. Chen et al.~\cite{chen2021badnl} first explore the effectiveness of backdoor attacks in natural language process, they proposed three different classes of triggers and achieved an almost perfect attack success rate with a negligible effect on the benign model's utility. 
%Sun et al.~\cite{sun2020natural} also propose natural triggers to attack NLP models with preserving the semantics of the sentence. 
Kurita~\cite{kurita2020weight} et al. propose a weighted poisoning attack named RIPPLe to attack the pre-trained models in fine-tuning phase. Their experiments on sentiment classification, toxicity detection, and spam detection show that RIPPLe is widely applicable and posed a serious threat. 
%Unlike RIPPLE, Shen et al.~\cite{shen2021backdoor} map the inputs containing triggers directly to a predefined output representation of the pre-trained NLP models, instead of a target label. Thus, the backdoor is introduced to downstream tasks without prior knowledge. Their method is also widely applicable to different fine-tuning tasks and different models which posed a severe threat.
Backdoor attacks have also been studied on code-related tasks~\cite{li2022poison} such as code completion~\cite{ramakrishnan2020backdoors} and code search~\cite{backdoorfse}. However, backdoor attacks on code search suffer from a low attack success rate and bad performance on infected models.
\subsection{Neural Code Search}
With the development of deep learning, the performance of neural code search has been significantly advanced~\cite{gu2018deep,cambronero2019deep,liu2020simplifying}. 
%The main idea of deep learning based neural code search is to map the natural language and the code into high-dimensional vectors and then train the models to match the natural language and corresponding code snippet according to their vectors' similarity. 
Gu et al.~\cite{gu2018deep} first propose DeepCS, which represents the method name, API sequence, and token sequence using the RNN and then measure the similarity of code and query in embedding space. %then computes the cosine similarity of them. 
%Shai et al.~\cite{shuai2020improving} also represent the method name, API sequence, and token sequence, but they then merge these features using a feature matrix. 
Wan et al.~\cite{wan2019multi} introduce MMAN which considers the structural features of code(e.g., Abstract Syntax Tree and control-flow graph) and use the attention mechanism to improve the effectiveness of MMAN. Ling et al.~\cite{wan2019multi} propose to convert the natural language and code into graphs using RGCN and strengthen the relationship between code and natural language using a semantic matching module. Sun et al.~\cite{sun2022code}, however, compile the code to instruction sequence and use the translation rules to translate the instruction sequence into its corresponding natural language as the representation of code. 
%Chai et al.~\cite{chai2022cross} proposed CDCS which uses meta-learning to resolve the problem that the domain-specific languages have relatively scarce data.
However, in this work, we attempt to attack code search models.
\section{Conclusion}
In this paper, we investigate backdoor attacks on deep learning based code search models and propose a novel method named \tool that can %to 
successfully attack code search models.
% while preserving or even achieving %getting 
% better performance compared with benign models. 
The experiments on four code search models and two datasets demonstrate the effectiveness of our method in attacking code search models. It also shows %proves 
the fragile of deep learning based code search models and the importance of research in security and robustness.
The experiments on two backdoor defense methods also demonstrate that existing backdoor defense methods cannot defend %defense 
against \tool effectively. In the future, we plan to explore more invisible trigger designs. We also plan to investigate effective defense strategies on backdoor attack for neural code models. 

\bibliographystyle{IEEEtran}
\bibliography{section_reference}

\end{document}